\documentclass[journal]{IEEEtran}

\usepackage{amsthm,amsfonts,amssymb,mathtools,nccmath,mathrsfs}
\usepackage[lined,linesnumbered,ruled,commentsnumbered]{algorithm2e}
\usepackage{pgfplots}
\usepackage{xcolor}
\usepackage{microtype}
\usepackage{url}
\usepackage{setspace}
\usepackage{array}
\usepackage{subfigure}
\usepackage{textcomp}
\usepackage{stfloats}
\usepackage{verbatim}
\usepackage{graphicx}
\usepackage{cite}
\usepackage{hyperref}
\usepackage{soul}
\usetikzlibrary{arrows,shapes,backgrounds,plotmarks,positioning}
\usepackage{tikz}
\usetikzlibrary{decorations.pathreplacing}
\usepackage{interval}

\newtheorem{definition}{Definition}
\newtheorem{corollary}{Corollary}
\newtheorem{proposition}{Proposition}

\newtheorem{example}{Example}
\newtheorem{remark}{Remark}

\newcommand{\X}{\mathcal{X}}
\newcommand{\Y}{\mathcal{Y}}
\newcommand{\Sen}{\mathcal{S}}
\newcommand{\M}{\mathcal{M}}

\newcommand{\PX}{P_{X}}
\newcommand{\PS}{P_{S}}
\newcommand{\PY}{P_{Y}}
\newcommand{\PSX}{P_{SX}}
\newcommand{\PSY}{P_{SY}}
\newcommand{\PXY}{P_{XY}}
\newcommand{\PXgY}{P_{X|Y}}

\newcommand{\PYgX}{P_{Y|X}}

\newcommand{\PSgX}{P_{S|X}}

\newcommand{\PSgY}{P_{S|Y}}

\newcommand{\e}{\mathrm{e}}
\newcommand{\eps}{\varepsilon}

\newcommand{\Real}{\mathbb{R}}

\begin{document}

\title{An Algorithm for Enhancing Privacy-Utility Tradeoff in the Privacy Funnel and Other Lift-based Measures}

\author{
    Mohammad A.~Zarrabian,~\IEEEmembership{Member,~IEEE,} and~Parastoo~Sadeghi,~\IEEEmembership{Senior Member,~IEEE}
    \thanks{Mohammad A.~Zarrabian is with the College of Engineering, Computing, and Cybernetics, Australian National University, Canberra, Australia, e-mail: mohammad.zarrabian@anu.edu.au. Parastoo Sadeghi is with the School of Engineering and Technology, the University of New South Wales, Canberra, Australia, e-mail: p.sadeghi@unsw.edu.au.}
}
\maketitle

\begin{abstract}
    This paper investigates the privacy funnel, a privacy-utility tradeoff problem in which mutual information quantifies both privacy and utility. 
    The objective is to maximize utility while adhering to a specified privacy budget. 
    However, the privacy funnel represents a non-convex optimization problem, making it challenging to achieve an optimal solution.
    An existing proposed approach to this problem involves substituting the mutual information with the lift (the exponent of information density) and then solving the optimization. Since mutual information is the expectation of the information density, this substitution overestimates the privacy loss and results in a final smaller bound on the privacy of mutual information than what is allowed in the budget. This significantly compromises the utility.

    To overcome this limitation, we propose using a privacy measure that is more relaxed than the lift but stricter than mutual information while still allowing the optimization to be efficiently solved. Instead of directly using information density, our proposed measure is the average of information density over the sensitive data distribution for each observed data realization.
    We then introduce a heuristic algorithm capable of achieving solutions that produce extreme privacy values, which enhances utility. The numerical results confirm improved utility at the same privacy budget compared to existing solutions in the literature.
    
    Additionally, we explore two other privacy measures, $\ell_{1}$-norm and strong $\chi^2$-divergence, demonstrating the applicability of our algorithm to these lift-based measures. We evaluate the performance of our method by comparing its output with previous works. Finally, we validate our heuristic approach with a theoretical framework that estimates the optimal utility for strong $\chi^2$-divergence, numerically showing a perfect match.
\end{abstract}


\section{Introduction}
    There has been a rapid development in machine learning and data analysis, as well as in communication networks such as 5G and 6G. This has led to a growing demand for data sharing over wireless networks. Consequently, the importance of data privacy and secrecy has become increasingly significant, as most data sets contain personal and sensitive information that must be protected \cite{2018FacebookGuardian,2021SecrecybyPF,2020PFInternetRobot}.
    
    To address this, various privacy measures have been developed to quantify the extent of sensitive information leakage \cite{2006DFP,2013LDPMiniMax,2020MaxL,2019TunMsurInfLeak_PUT,2021Alpha-LiftWatchdog,2021StrongChi2,saeidian2023pointwise,2023Onthelift,grosse2024quantifying}. One common strategy for protecting privacy is data randomization, where a privacy mechanism generates a randomized version of the dataset for publication rather than releasing the original data \cite{2014ExtermalMechanism,2019Watchdog,2020PropertiesWatchdog,2022EnhanceUtilWatchdog,2020PerfectObfusc,2022asymmetric}. 
    However, this randomization typically reduces data utility. In the extreme case of perfect privacy, where there is zero information leakage,  the released dataset is practically rendered useless \cite{2021OnPerfectPrivacy,saeidian2023inferential}. Therefore, some level of controlled privacy leakage is necessary to facilitate effective data sharing.
    Utility measures are defined to quantify data usefulness to strike a balance between privacy and data utility, and the desired privacy mechanism is derived from an optimization problem known as the privacy-utility tradeoff (PUT). This problem involves maximizing utility within a predetermined privacy leakage limit (the \textit{privacy budget}) or minimizing privacy leakage for a given utility level.

    In information-theoretic privacy, information leakage is often measured by examining changes in an adversary’s belief about sensitive features before and after observing the data. Various statistical information divergences, such as mutual information (MI)  \cite{2012PrivStatisticInfere,2013UPTdatasets,2014PFInfBottneck}, total variation distance \cite{2019PUTTotalDistnce}, $\chi^2$-divergence \cite{2019PrivEstimGuarant,2021StrongChi2}, Sibson and Arimoto MI \cite{2018TunableMsurInfleak,2023Generalgain} are commonly used as privacy measures. These measures quantify leakage as an expectation over the joint distribution of sensitive and observed data and are thus referred to as average measures \cite{2012PrivStatisticInfere,2021ContextawareLIP}.

    Beyond average measures, other categories have been introduced to provide stronger privacy guarantees, such as pointwise and semi-pointwise measures. Pointwise measures, include local differential privacy (LDP) \cite{2003LimitPrivBreach,2013LDPMiniMax,2014ExtermalMechanism}, local information privacy (LIP) \cite{2012PrivStatisticInfere,2021ContextawareLIP,grosse2024quantifying,2022asymmetric}, and pointwise maximal leakage \cite{saeidian2023pointwise,2023extremalPML} (also known as max-lift \cite{2023Onthelift,grosse2024quantifying}). They offer privacy guarantees for each individual realization of sensitive and observed data rather than an average value. Semi-pointwise measures, such as $\alpha$-lift \cite{2021Alpha-LiftWatchdog,2024AlphaAlgorithm}, $\ell_{1}$-norm \cite{2019DataDsclsurPtfPriv, 2023Onthelift}, and strong $\chi^2$-divergence \cite{2021StrongChi2,2023Onthelift}, are more relaxed compared to pointwise measures, but stronger than the average ones. They quantify privacy for each realization of observed data as an average over sensitive information.

     The privacy funnel \cite{2014PFInfBottneck,2018GeneralBottneckProb,2019SubmodClustrInfbottPF,2020BottneckInfEstim,2024PfConvexsolver,2021SecrecybyPF,2018GenrelPF} is the dual of the information bottleneck \cite{2000InfBottNeck,2020BottneckInfEstim,2023BottlCLub} and a PUT problem where MI is used to measure both privacy and utility. It is applied to privacy-preserving machine learning methods \cite{2024GraphPf,2022WiretapGenra,2024deepvaria,2024Deepfunnel}. 
     Since this problem is non-convex, finding an optimal solution is particularly challenging. Consequently, some previous works have proposed heuristic approaches based on merging methods \cite{2019SubmodClustrInfbottPF,2014PFInfBottneck}, although these can still degrade utility.
     In \cite{2021DataSanitize}, the authors replaced MI with lift (the exponent of information density) and focused on the LIP problem, which can ensure MI privacy. This approach transformed the problem into a convex optimization, allowing for an optimal solution that maximizes utility. 
     However, the privacy budget for the lift was set equal to that of MI, resulting in a final smaller bound on MI as an expected value of information density, ultimately leading to utility degradation.

    In this paper, we propose to use a semi-pointwise measure that is more relaxed than lift as a pointwise measure. Specifically, this measure represents the average of information density over the conditional distribution of sensitive data given each realization of observed data. We then apply this measure in place of MI as the privacy measure in the privacy funnel and propose an algorithm to enhance the PUT. Our algorithm is inspired by \cite{2024AlphaAlgorithm}, where a heuristic method was introduced to estimate the optimal utility when privacy was measured by $\alpha$-lift and utility by MI. 
    The core idea in  \cite{2024AlphaAlgorithm} involved raising the value of $\eps$ and then identifying the vertices of the polytope associated with maximum lift. After that, we selected the vertices that led to an extreme privacy measure ($\alpha$-lift in \cite{2024AlphaAlgorithm}) value.
    In the case of a convex problem, this approach provided a very good estimation of the optimal solution.  
    We show that our new algorithm can achieve a privacy mechanism that results in extreme values of the privacy measure. Our numerical experiments confirm its superiority in terms of utility at the same privacy budget compared to the best algorithm that existed in the literature before. 

    We also apply our method to other semi-pointwise measures such as $\ell_{1}$-norm and strong $\chi^2$-divergence and demonstrate the applicability of it. 
    Subsequently, we validate the efficiency of the algorithm by comparison to previous works. We demonstrate that using the semi-pointwise measure boosts utility compared to the LIP method in \cite{2021DataSanitize}, and also, our method outperforms the subset merging mechanism proposed in \cite{2023Onthelift}. Finally, we validate this heuristic approach to the theoretical framework for strong $\chi^2$-divergence and numerically show a perfect match.
 
    \subsection{Notation}

        All the random variables are discrete and defined on finite alphabets. 
        We use capital letters, e.g., $X$, to show random variables and lowercase letters, e.g., $x$, to denote their realizations.
        Uppercase calligraphic letters are applied to represent sets, such as $\X$ for the alphabet of $X$ with cardinality $|\X|$.
        Vectors and matrices are shown by bold uppercase letters, e.g., $\mathbf{W}=[W_{1},W_{2},\cdots,W_{n}]^{T}$.
        For random variables $X$ and $Y$, their joint probability distribution is shown by $\PXY$, and the probability of a specific event is given by  $\PXY(x,y)=\Pr[X=x,Y=y]$. Similarly, the marginal distribution is denoted by $\PX$, and the conditional probability of $X$ given a realization $Y=y$ is denoted by $\PXgY(x|y)$.

\section{System Models and Privacy Measures}

    Consider a Markov chain $S-X-Y$, where $X$ represents useful data intended to be shared and $S$ is a sensitive feature correlated with $X$ via $\PSX(s,x)\neq \PS(s)\PX(x)$. 
    To control privacy leakage, $Y$ is generated from $X$ via a privacy mechanism $\M$ given by the conditional distribution $\PYgX$ and shared, where $\PSgY(s|y)=\sum_{x}\PSgX(s|x)\PXgY(x|y)$ due to the Markov chain assumption.
    Privacy is quantified by a measure of information leakage between $S$ and $Y$. In this paper, we consider max-lift, mutual information, $\ell_{1}$-norm, and strong $\chi^2$-divergence\footnote{For the sake of brevity, we remove the prefix strong for $\chi^2$-divergence.}.
    The utility is also measured by mutual information as a popular candidate of sub-convex utility functions \cite{2014ExtermalMechanism,2023extremalPML}.
    The considered privacy measures are defined as follows:
    
    \begin{definition}
        For a joint distribution $\PSY$, \textbf{lift} is given by 
        \begin{align}
           l(s,y) = \frac{\PSgY(s|y)}{\PS(s)},
        \end{align}
        and \textbf{MI} is given as the expectation of log-lift (aka information density), $\log l(s,y)$,
        \begin{align}
            I(S;Y)=\sum_{s,y}\PSY(s,y)\log l(s,y).
        \end{align} 
        The \textbf{$\ell_{1}$-norm} is defined as  
        \begin{align}
            \ell_{1}(y) &= \sum_{s}\left|\PSgY(s|y)-\PS(s)\right| 
            = \sum_{s}\PS(s)\left|l(s,y)-1\right|, 
        \end{align}
        and \textbf{$\chi^{2}$-divergence} as:
        \begin{align}
               \chi^2(y) &= \sum_{s}\frac{\left(\PSgY(s|y)-\PS(s)\right)^{2}}{\PS(s)} 
                         = \sum_{s}\PS(s)\left( l(s,y)-1\right)^{2}.
        \end{align}

        \begin{remark}
            Note that the total variation distance is the expected value of $\ell_{1}(y)$
            \begin{align}
                T(S;Y)=\frac{1}{2}\sum_{y}\PY(y)\ell_{1}(y),
            \end{align}
            and the average $\chi^2$-divergence is given by 
            \begin{align}
                \chi^{2}(S;Y)=\sum_{y}\PY(y)\chi^2(y).
            \end{align}
        \end{remark}

      \begin{definition}\label{def:L(y)}
        We define another \textit{semi-pointwise} measure as 
        \begin{align}
            \mathfrak{L}(y)=\sum_{s}\PSgY(s|y)\log\frac{\PSgY(s|y)}{\PS(s)},
        \end{align}
        that is related to MI as follows:
        \begin{align}
            I(S;Y)=\sum_{y}\PY(y) \mathfrak{L}(y).
        \end{align}
    \end{definition}

    \begin{definition}
        Given an $\eps \in \Real_{+}$, a privacy mechanism $\M: \X \rightarrow \Y$ is called $\eps$-MI private w.r.t $S$ if: 
        \begin{align}
            I(S;Y) \leq \eps.
        \end{align}
        Similarly, $\eps$-$\ell_{1}$ and $\eps$-$\chi^2$ private mechanisms are respectively defined if
        \begin{align}
            \ell_{1}(y) \leq \eps, \quad \chi^2(y) \leq \eps^2.
        \end{align}
    \end{definition}

    \end{definition}    

     Since lift is the pointwise measure, bounding the max-lift $\max_{s,y}l(s,y)$ can imply a privacy guarantee for the average and semi-pointwise measures.
    This property will be utilized later in Algorithm \ref{alg:funnel}. This is captured and extended in the following proposition.
    
    \begin{proposition}\label{prop: liftbound}
        Given an $\eps\!\in\!\Real_{+}$, we have the following properties: 
         \begin{equation}
            \text{If} \max_{s,y}l(s,y) \leq \e^{\eps} \hspace{9pt} \Rightarrow \mathfrak{L}(y) \leq \eps,~I(S;Y) \leq \eps. \label{eq:lift-logavg}
            \end{equation}
            \begin{equation}
                 \text{If} ~\max_{s,y}l(s,y) \leq \eps+1   \Rightarrow \ell_{1}(y) \leq \eps,~T(S;Y) \leq \frac{\eps}{2}. \label{eq:lift-ell1}
            \end{equation}
           \begin{equation}
               \text{If} ~\max_{s,y}l(s,y) \leq \eps+1     \Rightarrow \chi^2(y) \leq \eps^2, ~\chi^2(S;Y) \leq \eps. \label{eq:lift-chi2}
           \end{equation}
    \end{proposition}
        
    Proposition \ref{prop: liftbound} implies that a privacy mechanism that satisfies a pointwise or semi-pointwise measure privacy guarantee also satisfies the corresponding average measure guarantee.  

    \begin{proposition}[\!\!{\cite{2019DataDsclsurPtfPriv}[Proposition 1]} \label{prop:zamaniprop}\!]
    The $\eps$-$\ell_{1}$ privacy criterion implies $\chi^{2}$-divergence privacy with different privacy budget, i.e.,  
        if $\ell_{1}(y) \leq \eps \Rightarrow \chi^{2}(y) \leq (\eps')^{2}$, where  $\eps'=\frac{\eps}{\sqrt{\min_{s}\PS(s)}}$. For the inverse direction,
        we have the following relationship: 
        \begin{align}
            \text {If}~~ \chi^2(y) \leq \eps^{2} \Rightarrow \ell_{1}(y) \leq \eps.
        \end{align}
    \end{proposition}
    \begin{corollary}[\!\!{\cite{2019DataDsclsurPtfPriv}[Corollary 1]} \label{corl:zamani}\!]
        Proposition \ref{prop:zamaniprop} indicates that for a given $\eps \in \Real_{+}$, an $\eps$-$\ell_{1}$ private mechanism results in higher utility than an $\eps$-$\chi^{2}$ private mechanism. 
    \end{corollary}

    \subsection{Privacy-utility tradeoff}

        In the privacy funnel, both privacy and utility are measured by MI, and the PUT optimization is given by:
        
    \begin{align}
        & \label{eq:PF} \max_{\substack{\PYgX}} I(X;Y) = H(X) - \min_{\substack{\PXgY,\PY}} H(X|Y)  \\ 
        \text{s.t.}~~ & \label{eq:PF-I(S,Y)}  I(S;Y) \leq \eps,  \\ 
        & \label{eq:PXgY cond}\sum\nolimits_{x}\PXgY(x|y)=1,~\PXgY(x|y) \geq 0,~ \forall x, y,  \\ 
        & \label{eq:Py cond}\sum\nolimits_{y}\PY(y)=1, ~\PY(y) \geq 0, \forall y,  \\ 
        & \label{eq:Py&PXgy cond}\sum\nolimits_{y}\PXgY(x|y)\PY(y)=\PX(x),~\forall x.
    \end{align}
    While we consider \eqref{eq:PF} in this paper, one can also consider the minimization of $I(S;Y)$ s.t. $I(X;Y)\geq r$.

    If we replace MI in \ref{eq:PF-I(S,Y)} with other privacy measures like the $\ell_{1}$-norm, we will have the corresponding PUT optimization. Therefore, in the rest of the paper, when discussing the PUT of any privacy measure, we will refer to \ref{eq:PF} together with the appropriate privacy measure under consideration.

    \subsection{Optimal max-lift mechanism}
    
        In \cite{2021DataSanitize}, \eqref{eq:lift-logavg} has been applied to achieve $\eps$-MI private mechanism in the privacy funnel via obtaining the optimal mechanism for max-lift privacy. 
        We review this method, which is the basis of our heuristic algorithm in Section \ref{sec:Hueristic}.
        
        For max-lift privacy, $I(S;Y)$ in \eqref{eq:PF-I(S,Y)} is replaced with 
        \begin{align}\label{eq:liftsycond}
             \frac{\PSgY(s|y)}{\PS(s)} \leq \e^{\eps}, ~~ \forall s,y \in \Sen \times \Y.
        \end{align}
        Using the Markov chain property, \eqref{eq:liftsycond} is expressed as a linear function of the main variable in \eqref{eq:PF}, $\PXgY(x|y)$, as follows:
        \begin{align}
            \label{eq:maxlift cond} \sum_{x}\PSgX(s|x)\PXgY(x|y) \leq \e^{\eps}\PS(s),~~ \forall s,y \in \Sen \times \Y,
        \end{align}
        which makes a convex polytope with other conditions \eqref{eq:PXgY cond}-\ref{eq:Py&PXgy cond}. 
        Due to the concavity of $H(X|Y)$  w.r.t $\PXgY$, the optimal solution exists as a linear combination of the vertices of the convex polytope derived from these conditions. 
        As a result, we can obtain the optimal mechanism via the following steps:   
        \begin{enumerate}
            \item \label{step:1}
            Let $\Delta_{\eps}$ denote the polytope of the privacy conditions for the max-lift given by:
            \begin{equation}\label{eq:polytope}
                \Delta_{\eps}=\left\{\hspace{-10pt}
                \begin{array}{ll}
                    &\mathbf{W} \in \mathbb{R}^{|\X|}: \\
                    &\displaystyle \sum_{x} W_{x}=1,~~  W_{x} \geq 0,~\forall x \in \X,\\
                    &\displaystyle \sum_{x}\frac{\PSgX(s|x)}{\PS(s)}W_{x} \leq \e^{\eps},~ \forall s \in \Sen.
                \end{array} \right\}
            \end{equation}
        
            \item \label{step:2}
            Obtain the vertices of $\Delta_{\eps}$ and denote by $\mathcal{W}_{\eps}=\{\mathbf{W}^{1}, \cdots, \mathbf{W}^M\}$  the set of all vertices. The  entropy of each $\mathbf{W}^{i}$ is shown by $h(\mathbf{W}^{i})$ and given by:
            $$h({\mathbf{W}}^{i})=-\sum_{x\in \X}W_{x}^{i}\log W_{x}^{i}.$$
            Subsequently, we can achieve the optimal $\PXgY$ and corresponding $\PY$ by solving the following linear program:
            \begin{equation}\label{eq:Py optimize}
                \begin{aligned}
                \min_{\mathbf{q} } & \sum\nolimits_{i=1}\nolimits^{M} q_{i}h(\mathbf{W}^{i}),\\
                \text{s.t.}~~ &\sum\nolimits_{i=1}\nolimits^{M} q_{i}=1,~q_{i}\geq 0,\\
                \quad & \sum\nolimits_{i=1}\nolimits^{M}q_{i}{W}^{i}_{x}=\PX(x) .
                \end{aligned}
            \end{equation}        
            \item \label{step:3}
            Let $\mathcal{T}=\{t: q_{t}>0 \}=\{\zeta_{1},\cdots,\zeta_{|\mathcal{T}|}\}$, which is the set of indices of nonzero elements in $\mathbf{q}$. Now, we have the sets $\Y$, $\PY$, and $\PXgY$ as follows:
            \begin{align}
                &\label{eq:output Y}\Y=\{1,\cdots,|\mathcal{T}|\},\\
                &\label{eq:output PY}\PY(y)=Q_{\zeta_{y}},~ y \in \Y,\\
                &\label{eq:output Pxgy}\PXgY(X|y)=\mathbf{W}^{\zeta_{y}},~ y \in \Y.
            \end{align}        
        \end{enumerate} 

        Although this approach archives the optimal solution for max-lift privacy, it still significantly degrades the privacy funnel's utility. 
        The reason is that  MI is the expected value of log-lift, and by abounding log-lift with $\eps$, which is the same privacy budget allocated for MI, we may end up with a much stricter bound on MI, leading to stricter privacy (than what is budgeted for) and low utility. 
        Therefore, we apply the semi-pointwise measure $\mathfrak{L}(y)$ to achieve a closer bound to $\eps$ for MI.
         In the upcoming section, we present our algorithm as an extension of the max-lift optimal solution to improve the privacy funnel, which we will later also use to estimate the optimal utility for the $\ell_{1}$-norm and $\chi^2$-divergence.

\section{Heuristic algorithm to enhance PUT \label{sec:Hueristic} }
    
    To address the issue mentioned above of the max-lift mechanism, we utilize $\mathfrak{L}(y)$ in Definition \ref{def:L(y)} instead of $I(S;Y)$ in \eqref{eq:PF-I(S,Y)} and we have:
    \begin{align}
         & \label{eq:opt L(y)} \max_{\substack{\PYgX}} I(X;Y) = H(X) - \min_{\substack{\PXgY,\PY}} H(X|Y)  \\ 
        \text{s.t.}~~ & \label{eq:cond L(y)} \hspace{10pt} \sum_{s}\PSgY(s|y)\log\frac{\PSgY(s|y)}{\PS(s)} \leq \eps.  
    \end{align}
    where $\PSgY(s|y)=\sum_{x}\PSgX(s|x)\PXgY(x|y)$ and other conditions \eqref{eq:PXgY cond}-\eqref{eq:Py&PXgy cond} are unchanged.
    
    The utility is known to be an increasing function of $\eps$ \cite{2020BottneckInfEstim}. Consequently, we aim to design a mechanism where $\mathfrak{L}(y)$ approaches $\eps$ as closely as possible. Since this measure is more relaxed, bounding $\mathfrak{L}(y)$ by $\eps$, is expected to bring the resulting bound on $I(S;Y)$ closer to $\eps$ compared to the max-lift mechanism.

      \begin{algorithm}[b]
        \textbf{Input}: $\PSX$, $\mathcal{E}=\{\eps_1, \ldots, \eps_{|\mathcal{E}|} \}$, $\mathcal{N}=\{ n_1, \ldots, n_{|\mathcal{E}|} \}$, $\mathcal{U}_{\eps_i} = \emptyset$ for $0 \leq i \leq |\mathcal{E}|$, $\delta$, $\eps_{|\mathcal{E}|+1}$.\\

        \textbf{Output}: $\Y_{i}$, $\PXgY$, $\PY$, and $I_{i}(X;Y)$ for  $1 \leq i\leq |\mathcal{E}|$.\\
        
        \textbf{Initiate:}  Let $\mathcal{E}'_i=\{\eps_{i}+\frac{k(\eps_{i+1}-\eps_{i})}{n_{i}},0 \leq k \leq n_{i}-1\}$ for $1\leq i \leq |\mathcal{E}|$, $\mathcal{E}' = \cup_{i=1}^{|\mathcal{E}|} \mathcal{E}'_i$, and for  $1\leq i \leq |\mathcal{E}|$ let $\mathcal{F}_{\eps_{i}}=\{\PXgY(X|y) \text{~in~}\eqref{eq:output Pxgy}\}$.\\
        Obtain $\mathcal{R_{\eps'}}$ for all $\eps' \in \mathcal{E}'$. \\
      
        \For{$i=1:|\mathcal{E}|$}{
            $\mathcal{F}_{\eps_{i}}\leftarrow 
            \mathcal{F}_{\eps_{i}}
            \cup\mathcal{U}_{\eps_{i-1}} $\\
            \For{$k=i:|\mathcal{E}|$}{
                \begin{align*}\hspace{-10pt}
                   \nonumber \mathcal{F}_{\eps_{i}}\leftarrow
                    \mathcal{F}_{\eps_{i}}&
                    \cup \{\mathbf{W}\in\mathcal{R}_{\eps'}:\eps' \in \mathcal{E}'_k\\
                    &(1-\delta){\eps_{i}}\leq \mathfrak{L}(y) \leq \eps_{i} \};
                \end{align*}
            }
            Let  $\mathcal{W}_{\eps_{i}} = \mathcal{F}_{\eps_{i}}$ and solve \eqref{eq:Py optimize}.
            Use \eqref{eq:output Y}-\eqref{eq:output Pxgy} to determine $\Y_{i}$, $\PXgY$, $\PY$, $I_{i}(X;Y)$. 
            Let $\mathcal{U}_{\eps_{i}}= \{ \PXgY(X|y) s\text{~obtained for~} \eps_{i} \text{~in \eqref{eq:output Pxgy}}  \}$;
        }
        
        \caption{Estimation of $\!I(X;Y)$ in Privacy Funnel \label{alg:funnel}}
    \end{algorithm}

    Let $\mathcal{F}_{\eps}$ represent the set of feasible points $\PXgY(X|y)$ that satisfy the inequality in \eqref{eq:cond L(y)} for a given  $\eps$
    and let $\mathcal{R}_{\eps}$ denote the set of extremal points $\PXgY(X|y)$ that satisfy the inequality in \eqref{eq:cond L(y)} with equality. 
    If the average lift over $S$ satisfies $\mathfrak{L}(y)=\eps$, its maximum must be greater than $\eps$. That is, $\max_{s,y}l(s,y)>\e^{\eps}$. This forms the core idea of our approach.
     Therefore, if $\mathbf{W}\in \mathcal{R}_{\eps}$ is an extreme point for $\mathfrak{L}(y)$, it should belong to a corresponding max-lift polytope $\Delta_{\eps'}$ in \eqref{eq:polytope} for some $\eps'> \eps$. This characteristic suggests a heuristic method for identifying some of the extreme points in \eqref{eq:cond L(y)}.
   
    To solve the optimization,we propose a heuristic approach outlined in Algorithm~\ref{alg:funnel}. The primary inputs to Algorithm~\ref{alg:funnel} are 
    \(\PSX\), which is assumed to be known based on the available data, and an ordered set  \(\mathcal{E}=\{\eps_1, \ldots, \eps_{|\mathcal{E}|}\}\), where $\eps_i$'s are increasing. Additional auxiliary parameters will be discussed later. 
    The output is a feasible solution for \(\PXgY\), \(\PY\), the corresponding output alphabet \(\mathcal{Y}_i\), and the utility \(I_i(X;Y)\) for each \(\eps_{i} \in \mathcal{E}\).

    Due to the properties of privacy funnel \cite{2020BottneckInfEstim}, we expect that utility value increases when $\eps$ increases.
    However, since our approach is heuristic, the utility output for $ \eps_{i}$ may be slightly lower than the utility for  $ \eps_{i'}$ where  $\eps_{i}>\eps_{i'}$. To mitigate this, we consider a range of $\eps$ values.

    The algorithm begins by initializing $\mathcal{F}_{\eps_{i}}$ using $\PXgY(X|y)$s values from \eqref{eq:output Pxgy} for each $\eps_{i} \in \mathcal{E}$ (line 3).  
    These vectors are feasible but not extremal for $\eps_{i}$. 
     However, they ensure that if a solution is feasible for $\eps_{i}$ it remains feasible for any  $\eps_{i}<\eps_{j}$.
     Thus in line 6, for each $\eps_{i}$, we combine the feasible set $\mathcal{F}_{\eps_{i}}$  with the previous solutions $\eps_{i-1}$ saved in $\mathcal{U}_{\eps_{i-1}}$. This ensures that if a solution is achievable for $\eps_{i-1}<\eps_{i}$, then it is also achievable for $\eps_{i}$. Note the input $\mathcal{U}_{\eps_{0}}=\emptyset$ makes line 6 valid even for $i= 1$.
    
    Once $\mathcal{F}_{\eps_{i}}$  is established, lines 7-11 add additional points as follows.
    The values in set $\mathcal{E}'_i$ are initialized as $\eps' = \eps_{i}+\frac{k(\eps_{i+1}-\eps_{i})}{n_{i}}, 0 \leq k \leq n_{i}-1$. 
    For each $\eps' \in \mathcal{E'}_j$, we compute  $\mathcal{R}_{\eps'}$, the vertices of $\Delta_{\eps'}$ as per \eqref{eq:polytope}, and include those elements $\mathbf{W} \in \mathcal{E}_{\eps'}$ where the corresponding $\mathfrak{L}(y)$ is within the range $[(1-\delta)\eps_{i}, \eps_{i}]$,  for some small estimation tolerance $\delta$. 
This is exactly the part of the algorithm where we try extreme points of max-lift polytope for larger values of $\eps'$ to find possible good candidates for extreme points of the semi-pointwise measure $\mathfrak{L}(y)$ for the $\eps< \eps'$ of interest.
    
    For the largest value $\eps_{|\mathcal{E}|}$, we use the auxiliary input $\eps_{|\mathcal{E}|+1}>\eps_{|\mathcal{E}|}$ and interpolate $n_{|\mathcal{E}|}$ points between them to estimate $\mathcal{R}_{\eps_{|\mathcal{E}|}}$. 
    
    Finally, with $\mathcal{F}_{\eps_{i}}$ prepared, line 12 involves solving, \eqref{eq:PF} for $\eps_{i}$ and compute the utility using the 
    derived $\PXgY$ and $\PY$ from \eqref{eq:output Y}-\eqref{eq:output Pxgy}. 
    The results are saved as $\mathcal{U}_{\eps_{i}}$ for use in subsequent steps. 
    In the following section, we evaluate the performance of this method through numerical analysis.

\section{Numerical Results}

  \begin{figure*}[]
            \centering   
            \subfigure[\label{fig:utilityfunnel} Normalized utility]{\scalebox{0.4}{\definecolor{mycolor1}{rgb}{1,0,0}%
\definecolor{mycolor2}{rgb}{0.2, 0.8, 0.2}%
\definecolor{mycolor3}{rgb}{0,0,1}%
\definecolor{mycolor4}{rgb}{0.53, 0.81, 0.98}
\begin{tikzpicture}

\begin{axis}[%
    width=4.521in,
    height=3.4in,
    scale only axis,
    xlabel style={font=\color{white!15!black}, at={(axis description cs:0.5,-0.01)}},
    every x tick label/.append style={font=\color{darkgray!60!black},font=\Large},
    xlabel={\Huge $\eps$},
    xmin=0,
    xmax=0.5,
    ymin=0.3,
    ymax=1,
    ylabel style={font=\color{white!15!black}, at={(axis description cs:-0.03,0.5)}},
    ylabel={{\huge ${\frac{I(X;Y)}{H(X)}}$}},
    every y tick label/.append style={font=\color{darkgray!60!black},font=\Large},
    axis background/.style={fill=white},
    xmajorgrids,
    ymajorgrids,
 legend style={at={(0.515,0.02)}, anchor=south west, legend cell align=left, align=left, draw=white!15!black,font=\fontsize{19}{20}\selectfont}
    ]

\addplot [color=mycolor1, line width=2.0pt,
mark=square, mark options={scale=1.5, line width=0.8pt}]
  table[row sep=crcr]{%
0.005	0.569414265260603\\
0.01	0.630917691268975\\
0.015	0.674773759986823\\
0.02	0.705176422524871\\
0.025	0.72792035126865\\
0.03	0.746909140150921\\
0.035	0.763487855649957\\
0.04	0.77889020925859\\
0.045	0.793139039480012\\
0.05	0.805820673427334\\
0.055	0.818181419478711\\
0.06	0.828475752913862\\
0.065	0.838020737464444\\
0.07	0.848000472581122\\
0.075	0.85609281042734\\
0.08	0.863910668993627\\
0.085	0.870858389534913\\
0.09	0.877652507290172\\
0.095	0.883860470908083\\
0.1	0.889830826261929\\
0.105	0.895662667521707\\
0.11	0.901251829915382\\
0.115	0.906331646799496\\
0.12	0.91114901713155\\
0.125	0.915548220431229\\
0.13	0.919629277617068\\
0.135	0.923770355981742\\
0.14	0.9273824274492\\
0.145	0.93097203139187\\
0.15	0.934386907450872\\
0.155	0.937742761446795\\
0.16	0.940898130219757\\
0.165	0.944030693327656\\
0.17	0.947094286643168\\
0.175	0.950023392892225\\
0.18	0.952931791325715\\
0.185	0.955715816624921\\
0.19	0.958306453993227\\
0.195	0.960519611173797\\
0.2	    0.96266188733157\\
0.205	0.96459764128295\\
0.21	0.966575691516578\\
0.215	0.968457302078595\\
0.22	0.970070447693458\\
0.225	0.971531338863279\\
0.23	0.972970731533955\\
0.235	0.97438558498429\\
0.24	0.975720469862802\\
0.245	0.97692534209513\\
0.25	0.977976680990648\\
0.255	0.978886362228259\\
0.26	0.97982119758694\\
0.265	0.980664310549639\\
0.27	0.981401790879489\\
0.275	0.982100847162786\\
0.28	0.982811091173999\\
0.285	0.983472564540472\\
0.29	0.984135883823792\\
0.295	0.984736088706078\\
0.3	0.985361740675461\\
0.305	0.985955042326312\\
0.31	0.986583245241448\\
0.315	0.98722278831089\\
0.32	0.987823193878181\\
0.325	0.988342175751448\\
0.33	0.988773064783533\\
0.335	0.989161404163436\\
0.34	0.989588265392407\\
0.345	0.989969551976176\\
0.35	0.990291390342828\\
0.355	0.990606967844951\\
0.36	0.990865334260461\\
0.365	0.991124437471994\\
0.37	0.991328428368992\\
0.375	0.991514557891293\\
0.38	0.991687606047694\\
0.385	0.991850646695087\\
0.39	0.991998200701647\\
0.395	0.992140478579817\\
0.4	0.992272093091278\\
0.405	0.992426929634962\\
0.41	0.992563161912089\\
0.415	0.992697630935969\\
0.42	0.9927961653817\\
0.425	0.992900965283896\\
0.43	0.992981048902348\\
0.435	0.993054933120586\\
0.44	0.993116554265685\\
0.445	0.993181097449367\\
0.45	0.993238735822559\\
0.455	0.993293990902647\\
0.46	0.993334039298128\\
0.465	0.993360804601498\\
0.47	0.99339231784024\\
0.475	0.993425380202458\\
0.48	0.99345367019069\\
0.485	0.993482510289033\\
0.49	0.993511968384522\\
0.495	0.993541011927665\\
0.5	0.993572008399019\\
};
\addlegendentry{Algorithm 1}

\addplot [color=mycolor2, line width=2.0pt, mark=x, mark options={scale=1.5, line width=0.8pt}]
  table[row sep=crcr]{%
0.005	0.350867672899433\\
0.01	0.481159701064172\\
0.015	0.550451803653594\\
0.02	0.618068145555093\\
0.025	0.639373832492096\\
0.03	0.665189887919972\\
0.035	0.687398818349508\\
0.04	0.710131259357739\\
0.045	0.72248833452941\\
0.05	0.750794361042288\\
0.055	0.773674274709091\\
0.06	0.787361155511533\\
0.065	0.802838241306383\\
0.07	0.815972731695026\\
0.075	0.825007926372081\\
0.08	0.841437305633193\\
0.085	0.843487191510355\\
0.09	0.856340057621762\\
0.095	0.862327447027104\\
0.1	0.864789802799112\\
0.105	0.872936483879731\\
0.11	0.896697068861668\\
0.115	0.901341555160235\\
0.12	0.905540048797251\\
0.125	0.908743030033979\\
0.13	0.912755533038264\\
0.135	0.919876539096207\\
0.14	0.921586700239323\\
0.145	0.92669019779293\\
0.15	0.930093494491828\\
0.155	0.932612117130411\\
0.16	0.940112961196527\\
0.165	0.941619446706416\\
0.17	0.944522325332052\\
0.175	0.949831445582377\\
0.18	0.952081151444453\\
0.185	0.959900997460784\\
0.19	0.963832785636356\\
0.195	0.966992215203045\\
0.2	    0.970628021752501\\
0.205	0.973251243047285\\
0.21	0.973251243047285\\
0.215	0.974859282410702\\
0.22	0.979255634085206\\
0.225	0.979199993852771\\
0.23	0.978988615698713\\
0.235	0.978569338447557\\
0.24	0.981742834840944\\
0.245	0.9868815005665\\
0.25	0.9868815005665\\
0.255	0.986615899691313\\
0.26	0.987864006340375\\
0.265	0.987864006340375\\
0.27	0.987864006340375\\
0.275	0.987864006340375\\
0.28	0.987864006340375\\
0.285	0.987864006340375\\
0.29	0.989693412294383\\
0.295	0.989693412294383\\
0.3	0.989693412294383\\
0.305	0.989693412294383\\
0.31	0.989693412294383\\
0.315	0.99145045616522\\
0.32	0.991777441391077\\
0.325	0.991777441391077\\
0.33	0.991777441391077\\
0.335	0.991777441391077\\
0.34	0.993748429298327\\
0.345	0.993748429298327\\
0.35	0.993748429298327\\
0.355	0.995270316609739\\
0.36	0.995270316609739\\
0.365	0.995270316609739\\
0.37	0.995270316609739\\
0.375	0.995270316609739\\
0.38	0.995270316609739\\
0.385	0.995270316609739\\
0.39	0.996358320855912\\
0.395	0.996358320855912\\
0.4	0.996358320855912\\
0.405	0.996358320855912\\
0.41	0.996358320855912\\
0.415	0.998400658940183\\
0.42	0.998400658940183\\
0.425	0.998400658940183\\
0.43	0.998400658940183\\
0.435	0.998456299172618\\
0.44	0.998456299172618\\
0.445	0.998456299172618\\
0.45	0.998456299172618\\
0.455	0.998456299172618\\
0.46	0.998456299172618\\
0.465	0.998456299172618\\
0.47	0.998456299172618\\
0.475	0.998456299172618\\
0.48	0.998456299172618\\
0.485	0.998456299172618\\
0.49	0.998456299172618\\
0.495	0.998456299172618\\
0.5	0.998456299172618\\
};
\addlegendentry{Subset merging}

\addplot [color=mycolor3, line width=2.0pt, mark=o, mark options={scale=1.5, line width=0.8pt}]
  table[row sep=crcr]{%
0.005	0.40790849773531\\
0.01	0.420836321115586\\
0.015	0.43426942596536\\
0.02	0.446722337220316\\
0.025	0.458519166519926\\
0.03	0.470063444100365\\
0.035	0.481298074456325\\
0.04	0.492806707631815\\
0.045	0.503567394557631\\
0.05	0.513545312407512\\
0.055	0.523678230567413\\
0.06	0.533297889376264\\
0.065	0.542760060652131\\
0.07	0.551272785201024\\
0.075	0.55983871493695\\
0.08	0.568027292791412\\
0.085	0.57605615565116\\
0.09	0.583880574585311\\
0.095	0.59141865729959\\
0.1	0.599069901614855\\
0.105	0.606263538474519\\
0.11	0.613391629598037\\
0.115	0.620410794440911\\
0.12	0.62721750788812\\
0.125	0.634041151091353\\
0.13	0.640706751139674\\
0.135	0.647199607941321\\
0.14	0.653469189654325\\
0.145	0.659837291856392\\
0.15	0.666218620162139\\
0.155	0.672463874554373\\
0.16	0.678506147044804\\
0.165	0.684647503303312\\
0.17	0.690608336553258\\
0.175	0.696188677573964\\
0.18	0.701814235301762\\
0.185	0.707328108546786\\
0.19	0.712797103489877\\
0.195	0.718060059757151\\
0.2	0.72310300912433\\
0.205	0.727943979774884\\
0.21	0.732576320761532\\
0.215	0.737092794962517\\
0.22	0.741309701457259\\
0.225	0.745660782529247\\
0.23	0.749869796521261\\
0.235	0.754073924661263\\
0.24	0.758271609915809\\
0.245	0.762724801109569\\
0.25	0.766935182748346\\
0.255	0.77114133662624\\
0.26	0.775228639772323\\
0.265	0.779373404470381\\
0.27	0.783631689759919\\
0.275	0.787736549310748\\
0.28	0.791581860366201\\
0.285	0.795242870424933\\
0.29	0.7985344832774\\
0.295	0.801776575033124\\
0.3	0.805092130348293\\
0.305	0.808674124308206\\
0.31	0.812381878936404\\
0.315	0.816092096140005\\
0.32	0.819745927857733\\
0.325	0.823475838699215\\
0.33	0.826868557332295\\
0.335	0.830146691768723\\
0.34	0.83335000670963\\
0.345	0.836652886173095\\
0.35	0.839860089471585\\
0.355	0.843035040125638\\
0.36	0.846088360993916\\
0.365	0.84912962279697\\
0.37	0.852142275993232\\
0.375	0.855107136794119\\
0.38	0.858014857306755\\
0.385	0.860910194226626\\
0.39	0.863821095759757\\
0.395	0.866794824787868\\
0.4	0.869725676550183\\
0.405	0.872664731900154\\
0.41	0.875573415970192\\
0.415	0.878517697472283\\
0.42	0.881312722529383\\
0.425	0.884020581576693\\
0.43	0.886708952522351\\
0.435	0.889316734530456\\
0.44	0.891842412144973\\
0.445	0.894361441990482\\
0.45	0.896828430929318\\
0.455	0.899224881874496\\
0.46	0.901685458313649\\
0.465	0.904205392625799\\
0.47	0.906706496343437\\
0.475	0.909179891792961\\
0.48	0.911620579647858\\
0.485	0.914005807248749\\
0.49	0.916330559783302\\
0.495	0.918577111660529\\
0.5	0.920876542394806\\
};
\addlegendentry{Max-lift privacy}

\end{axis}
\end{tikzpicture}%
}}
            \subfigure[\label{fig:funnelleak} Privacy leakage]{\scalebox{0.4}{\definecolor{mycolor1}{rgb}{1,0,0}%
\definecolor{mycolor2}{rgb}{0.2, 0.8, 0.2}%
\definecolor{mycolor3}{rgb}{0,0,1}%
\definecolor{mycolor4}{rgb}{0.53, 0.81, 0.98}

\begin{tikzpicture}

\begin{axis}[%
width=4.521in,
height=3.4in,
scale only axis,
xlabel style={font=\color{white!15!black}},
every x tick label/.append style={font=\color{darkgray!60!black},font=\Large},
xlabel={\Huge $\eps$},
xmin=0,
xmax=0.5,
ymin=0,
ymax=0.14,
ylabel style={font=\color{white!15!black}, at={(axis description cs:-0.09,0.5)}},
ylabel={\huge $I(S,Y)$},
every y tick label/.append style={font=\color{darkgray!60!black},font=\Large},
xmajorgrids,
ymajorgrids,
legend style={at={(0.515,0.02)}, anchor=south west, legend cell align=left, align=left, draw=white!15!black,font=\fontsize{19}{20}\selectfont}
]
\addplot [color=mycolor1, line width=2.0pt,
mark=square, mark options={scale=1.5, line width=0.8pt}
]
  table[row sep=crcr]{%
0.005	0.00493160143937332\\
0.01	0.00979092809813588\\
0.015	0.0144813914559696\\
0.02	0.019013930586581\\
0.025	0.0233498635001218\\
0.03	0.0275303233356514\\
0.035	0.0315810269206719\\
0.04	0.0355948901550372\\
0.045	0.0393239907520155\\
0.05	0.0430220258507234\\
0.055	0.0466309500814566\\
0.06	0.0500186228178913\\
0.065	0.0532535041483848\\
0.07	0.0564397936681988\\
0.075	0.0595799962914304\\
0.08	0.062505846020014\\
0.085	0.0654232404084884\\
0.09	0.068178317965569\\
0.095	0.0708272151637149\\
0.1	0.0734685615758139\\
0.105	0.0759679355264382\\
0.11	0.0783815993758365\\
0.115	0.080634167633417\\
0.12	0.0828688625450772\\
0.125	0.0849450079484196\\
0.13	0.0869348036747033\\
0.135	0.0889094265684004\\
0.14	0.0907361023964975\\
0.145	0.0925436419076197\\
0.15	0.0943482168573048\\
0.155	0.0960272307124845\\
0.16	0.0976515911693327\\
0.165	0.0992763665642642\\
0.17	0.100877832207622\\
0.175	0.102456069663701\\
0.18	0.103929085677929\\
0.185	0.105324907602199\\
0.19	0.106668794377033\\
0.195	0.107872348794383\\
0.2	0.109028488434556\\
0.205	0.110118024414578\\
0.21	0.111195667947586\\
0.215	0.112233079654077\\
0.22	0.113160541814614\\
0.225	0.11402296334018\\
0.23	0.114873686779967\\
0.235	0.115735160315192\\
0.24	0.116452577426496\\
0.245	0.11717312510022\\
0.25	0.11778997206252\\
0.255	0.118367776733364\\
0.26	0.118971217441284\\
0.265	0.119497438408201\\
0.27	0.119987148754668\\
0.275	0.120447250379522\\
0.28	0.120921699414215\\
0.285	0.121366700345675\\
0.29	0.121808548287202\\
0.295	0.122202029760076\\
0.3	0.122629025435974\\
0.305	0.123036780399398\\
0.31	0.123417860607629\\
0.315	0.123811940571648\\
0.32	0.124174618421954\\
0.325	0.124502457558965\\
0.33	0.124792684457132\\
0.335	0.12505061141497\\
0.34	0.125333322912575\\
0.345	0.125578211766129\\
0.35	0.125770142587081\\
0.355	0.125978760818137\\
0.36	0.126159153823501\\
0.365	0.126310662236417\\
0.37	0.126454219019994\\
0.375	0.126577995565904\\
0.38	0.126699989472953\\
0.385	0.126802099108536\\
0.39	0.126915027949568\\
0.395	0.127007054280748\\
0.4	0.127108272580637\\
0.405	0.127210606624952\\
0.41	0.127289960934177\\
0.415	0.127377069082232\\
0.42	0.127437864245863\\
0.425	0.12751057875488\\
0.43	0.127568809559308\\
0.435	0.12761565039352\\
0.44	0.12766355741756\\
0.445	0.127708928424441\\
0.45	0.127746629469257\\
0.455	0.127786880491118\\
0.46	0.127817780695971\\
0.465	0.127841846738024\\
0.47	0.127869684285188\\
0.475	0.127899236824403\\
0.48	0.127924236671319\\
0.485	0.127949547616757\\
0.49	0.127975192087746\\
0.495	0.12799992128893\\
0.5	0.128026305062256\\
};
\addlegendentry{Algorithm 1}

\addplot [color=mycolor2, line width=2.0pt, mark=x, mark options={scale=1.5, line width=0.8pt}]
  table[row sep=crcr]{%
0.005	0.00202129340867432\\
0.01	0.00715873579660502\\
0.015	0.0113174605635021\\
0.02	0.0195014431625522\\
0.025	0.0210297697779015\\
0.03	0.0242686921697342\\
0.035	0.0269422477823239\\
0.04	0.0304185398551965\\
0.045	0.0323241381819127\\
0.05	0.0383259846008172\\
0.055	0.0436811206723919\\
0.06	0.0487374950218427\\
0.065	0.0533863814955338\\
0.07	0.0566485183017505\\
0.075	0.0584942362719028\\
0.08	0.0618073831157247\\
0.085	0.0613560170921527\\
0.09	0.0653676156189593\\
0.095	0.0681861766875163\\
0.1	0.0682394957935645\\
0.105	0.071126807598368\\
0.11	0.0774744820452898\\
0.115	0.0797334483983115\\
0.12	0.0820341451873952\\
0.125	0.0843039780955602\\
0.13	0.0853222489043406\\
0.135	0.0886271694845452\\
0.14	0.0893730532438482\\
0.145	0.0915828531886371\\
0.15	0.0924225710098444\\
0.155	0.0937523803444892\\
0.16	0.0980151175186479\\
0.165	0.0980961115878878\\
0.17	0.0986840221407445\\
0.175	0.101311332485692\\
0.18	0.103633616560752\\
0.185	0.107476492967017\\
0.19	0.109479075967562\\
0.195	0.110477949563867\\
0.2	0.112404432063589\\
0.205	0.113202926948111\\
0.21	0.113202926948111\\
0.215	0.113377360211167\\
0.22	0.114812050789404\\
0.225	0.115149474071643\\
0.23	0.114870348169599\\
0.235	0.114889669553909\\
0.24	0.116894572876003\\
0.245	0.119952969175139\\
0.25	0.119952969175139\\
0.255	0.120822032615396\\
0.26	0.121825979724324\\
0.265	0.121825979724324\\
0.27	0.121825979724324\\
0.275	0.121825979724324\\
0.28	0.121825979724324\\
0.285	0.121825979724324\\
0.29	0.12339516370452\\
0.295	0.12339516370452\\
0.3	0.12339516370452\\
0.305	0.12339516370452\\
0.31	0.12339516370452\\
0.315	0.124274597589982\\
0.32	0.124537348157693\\
0.325	0.124537348157693\\
0.33	0.124537348157693\\
0.335	0.124537348157693\\
0.34	0.125040973267483\\
0.345	0.125040973267483\\
0.35	0.125040973267483\\
0.355	0.126048363456993\\
0.36	0.126048363456993\\
0.365	0.126048363456993\\
0.37	0.126048363456993\\
0.375	0.126048363456993\\
0.38	0.126048363456993\\
0.385	0.126048363456993\\
0.39	0.12720905538109\\
0.395	0.12720905538109\\
0.4	0.12720905538109\\
0.405	0.12720905538109\\
0.41	0.12720905538109\\
0.415	0.128313364006461\\
0.42	0.128313364006461\\
0.425	0.128313364006461\\
0.43	0.128313364006461\\
0.435	0.127975940724222\\
0.44	0.127975940724222\\
0.445	0.127975940724222\\
0.45	0.127975940724222\\
0.455	0.127975940724222\\
0.46	0.127975940724222\\
0.465	0.127975940724222\\
0.47	0.127975940724222\\
0.475	0.127975940724222\\
0.48	0.127975940724222\\
0.485	0.127975940724222\\
0.49	0.127975940724222\\
0.495	0.127975940724222\\
0.5	0.127975940724222\\
};
\addlegendentry{Subset merging}

\addplot [color=mycolor3, line width=2.0pt, mark=o, mark options={scale=1.5, line width=0.8pt}]
  table[row sep=crcr]{%
0.005	3.53707621103729e-05\\
0.01	0.000134824788692457\\
0.015	0.000286279491922503\\
0.02	0.000480653877078049\\
0.025	0.000711346776229234\\
0.03	0.000997603050298859\\
0.035	0.00132354679275826\\
0.04	0.00168039708753565\\
0.045	0.00204116024925426\\
0.05	0.00247004314879404\\
0.055	0.00295320810183096\\
0.06	0.00338733018497092\\
0.065	0.00391056928628766\\
0.07	0.00448825386623057\\
0.075	0.00499756415356805\\
0.08	0.00556779249634131\\
0.085	0.0062528778309325\\
0.09	0.00688643050215964\\
0.095	0.00760863192296682\\
0.1	0.00835345823011768\\
0.105	0.0089726448246913\\
0.11	0.00965756730600071\\
0.115	0.0102909144412573\\
0.12	0.0110970413349239\\
0.125	0.0118070526992211\\
0.13	0.0125894601637286\\
0.135	0.0135209163556838\\
0.14	0.0143996823629606\\
0.145	0.0154107064591272\\
0.15	0.0163787323092666\\
0.155	0.0172242415707799\\
0.16	0.0182296715547833\\
0.165	0.0194051756554967\\
0.17	0.0202718523719843\\
0.175	0.0215047208457915\\
0.18	0.0224285764784009\\
0.185	0.0235154658225185\\
0.19	0.0246412190813312\\
0.195	0.0256386963809379\\
0.2	0.0269898844893925\\
0.205	0.0279316187339312\\
0.21	0.0288370463575977\\
0.215	0.0298859118959158\\
0.22	0.0308139505683144\\
0.225	0.0317220077140044\\
0.23	0.0327215108592462\\
0.235	0.0335467660021089\\
0.24	0.0346267753650378\\
0.245	0.0357871779812187\\
0.25	0.036921981431512\\
0.255	0.0380083445123924\\
0.26	0.0391951934570334\\
0.265	0.0410214259897204\\
0.27	0.0422596425084884\\
0.275	0.0435152408839868\\
0.28	0.0448294405434396\\
0.285	0.0457014458315206\\
0.29	0.0467086734115891\\
0.295	0.0478685249469326\\
0.3	0.0488359555033076\\
0.305	0.0500368685209009\\
0.31	0.0510080147371676\\
0.315	0.0519917399202532\\
0.32	0.0530301736638418\\
0.325	0.0545232482554561\\
0.33	0.0554758461681621\\
0.335	0.0566754337176045\\
0.34	0.0583487317953096\\
0.345	0.0592594764851051\\
0.35	0.0603121462619496\\
0.355	0.0614807411005242\\
0.36	0.0627428885334802\\
0.365	0.0636811155708009\\
0.37	0.0647264405527311\\
0.375	0.0661885068900098\\
0.38	0.0672002527933467\\
0.385	0.0681054882012471\\
0.39	0.0690150438897273\\
0.395	0.0703580984054201\\
0.4	0.071152477332798\\
0.405	0.07251774710708\\
0.41	0.0734471069572628\\
0.415	0.0745621721583452\\
0.42	0.0753634028447559\\
0.425	0.0763479504000153\\
0.43	0.0771188379874522\\
0.435	0.0779990732322094\\
0.44	0.0787689501762468\\
0.445	0.0797302846222743\\
0.45	0.0806954557613255\\
0.455	0.0816566622918312\\
0.46	0.0824976661303126\\
0.465	0.0834137442666977\\
0.47	0.0844691423912027\\
0.475	0.085471565074962\\
0.48	0.0868352591135617\\
0.485	0.0878135404764526\\
0.49	0.0887546459519888\\
0.495	0.0897107694382121\\
0.5	0.0906579374294092\\
};
\addlegendentry{Max-lift privacy}
]

\end{axis}

\end{tikzpicture}%
}}
            \subfigure[\label{fig:funnelmax} Max-lift leakage]{\scalebox{0.4}{\definecolor{mycolor1}{rgb}{1,0,0}%
\definecolor{mycolor2}{rgb}{0.2, 0.8, 0.2}%
\definecolor{mycolor3}{rgb}{0,0,1}%
\definecolor{mycolor4}{rgb}{0.53, 0.81, 0.98}

\begin{tikzpicture}

\begin{axis}[%
width=4.521in,
height=3.4in,
scale only axis,
xlabel style={font=\color{white!15!black}},
every x tick label/.append style={font=\color{darkgray!60!black},font=\Large},
xlabel={\Huge $\eps$},
xmin=0,
xmax=0.5,
ymin=0,
ymax=0.8,
ylabel style={font=\color{white!15!black}, at={(axis description cs:-0.03,0.5)}},
ylabel={\huge $\max_{s,y}l(s,y)$},
every y tick label/.append style={font=\color{darkgray!60!black},font=\Large},
axis background/.style={fill=white},
xmajorgrids,
ymajorgrids,
legend style={at={(0.515,0.02)}, anchor=south west, legend cell align=left, align=left, draw=white!15!black,font=\fontsize{19}{20}\selectfont}
]
\addplot [color=mycolor1, line width=2.0pt,
mark=square, mark options={scale=1.5, line width=0.8pt}
]
  table[row sep=crcr]{%
0.005	0.137379999999999\\
0.01	0.205724444011513\\
0.015	0.257717801285283\\
0.02	0.300202142073451\\
0.025	0.329639999999999\\
0.03	0.358746334212892\\
0.035	0.381024183016269\\
0.04	0.403633022394003\\
0.045	0.424666287880353\\
0.05	0.447961672153034\\
0.055	0.465921003259615\\
0.06	0.479526302054012\\
0.065	0.493569646657108\\
0.07	0.505891091552839\\
0.075	0.517241486145862\\
0.08	0.52970698195721\\
0.085	0.540680327222665\\
0.09	0.54972476028145\\
0.095	0.561138027230792\\
0.1	0.570758027230792\\
0.105	0.579798027230792\\
0.11	0.588350470772952\\
0.115	0.596353120725617\\
0.12	0.603798120403524\\
0.125	0.611006559748222\\
0.13	0.620546560673092\\
0.135	0.627105039801838\\
0.14	0.633753615456859\\
0.145	0.640340839212297\\
0.15	0.646660839212297\\
0.155	0.655300047057255\\
0.16	0.661300113844972\\
0.165	0.667160113844972\\
0.17	0.67269453352411\\
0.175	0.67821453352411\\
0.18	0.683469533846202\\
0.185	0.688313429283122\\
0.19	0.693002685672696\\
0.195	0.697130466363494\\
0.2	0.701170466363494\\
0.205	0.703927489612464\\
0.21	0.707632142523576\\
0.215	0.710998005078961\\
0.22	0.713883640663732\\
0.225	0.716503640663732\\
0.23	0.719198689321654\\
0.235	0.721466387202904\\
0.24	0.723484205305447\\
0.245	0.725514795823523\\
0.25	0.727213397772194\\
0.255	0.728880635535631\\
0.26	0.730483612286661\\
0.265	0.732069973013798\\
0.27	0.73379225771455\\
0.275	0.735182647377142\\
0.28	0.736762647377142\\
0.285	0.738162647377142\\
0.29	0.739622647377142\\
0.295	0.740936286650006\\
0.3	0.742456286650006\\
0.305	0.743863892824303\\
0.31	0.745343892824303\\
0.315	0.746795972275176\\
0.32	0.748293898842506\\
0.325	0.749761640723614\\
0.33	0.750721640723614\\
0.335	0.751661640723614\\
0.34	0.752681640723614\\
0.345	0.753538143005307\\
0.35	0.754238143005307\\
0.355	0.754993937811936\\
0.36	0.755673937811936\\
0.365	0.756304078342087\\
0.37	0.756884078342087\\
0.375	0.757451916935243\\
0.38	0.757951916935243\\
0.385	0.758451916935242\\
0.39	0.758887219608234\\
0.395	0.759327219608234\\
0.4	0.759727219608235\\
0.405	0.760107219608234\\
0.41	0.760427219608234\\
0.415	0.760748697615689\\
0.42	0.760988697615689\\
0.425	0.761308697615689\\
0.43	0.761548697615689\\
0.435	0.761732995935702\\
0.44	0.761972995935702\\
0.445	0.762172995935702\\
0.45	0.762368697615689\\
0.455	0.762648697615689\\
0.46	0.762868220867939\\
0.465	0.763048220867939\\
0.47	0.763248220867939\\
0.475	0.763468220867939\\
0.48	0.763648220867939\\
0.485	0.763828220867939\\
0.49	0.764008220867939\\
0.495	0.764168220867939\\
0.5	0.764348220867939\\
};
\addlegendentry{Algorithm 1}

\addplot  [color=mycolor2, line width=2.0pt, mark=x, mark options={scale=1.5, line width=0.8pt}]
  table[row sep=crcr]{%
0.005	0.0993495014154188\\
0.01	0.179155716386961\\
0.015	0.239382843676177\\
0.02	0.334391376280148\\
0.025	0.348318595691259\\
0.03	0.363939799455503\\
0.035	0.408350715406755\\
0.04	0.434753902201542\\
0.045	0.456585806826468\\
0.05	0.487620333488757\\
0.055	0.509237225804394\\
0.06	0.533121795261424\\
0.065	0.550008381849531\\
0.07	0.557050419129274\\
0.075	0.570202170556224\\
0.08	0.578426059445342\\
0.085	0.587075957045919\\
0.09	0.603053818157611\\
0.095	0.608964560871856\\
0.1	0.608964560871856\\
0.105	0.606990982302907\\
0.11	0.619326399252042\\
0.115	0.625896808990252\\
0.12	0.629087217536287\\
0.125	0.644128693302135\\
0.13	0.651145017044961\\
0.135	0.673442082846838\\
0.14	0.673442082846838\\
0.145	0.681029754433356\\
0.15	0.683419721986621\\
0.155	0.683419721986621\\
0.16	0.694320352814429\\
0.165	0.694320352814429\\
0.17	0.696249315224756\\
0.175	0.705443598480514\\
0.18	0.7142708785839\\
0.185	0.715885542930897\\
0.19	0.71866799575647\\
0.195	0.724553889088878\\
0.2	0.723058408318639\\
0.205	0.724553889088878\\
0.21	0.724553889088878\\
0.215	0.725004751317948\\
0.22	0.730679725957664\\
0.225	0.723658986190864\\
0.23	0.724680493585465\\
0.235	0.724680493585465\\
0.24	0.735436361023278\\
0.245	0.747380388954865\\
0.25	0.747380388954865\\
0.255	0.747380388954865\\
0.26	0.750274811444\\
0.265	0.750274811444\\
0.27	0.750274811444\\
0.275	0.750274811444\\
0.28	0.750274811444\\
0.285	0.750274811444\\
0.29	0.752997627582517\\
0.295	0.752997627582517\\
0.3	0.752997627582517\\
0.305	0.752997627582517\\
0.31	0.752997627582517\\
0.315	0.752997627582517\\
0.32	0.753799798520829\\
0.325	0.753799798520829\\
0.33	0.753799798520829\\
0.335	0.753799798520829\\
0.34	0.753799798520829\\
0.345	0.753799798520829\\
0.35	0.753799798520829\\
0.355	0.753799798520829\\
0.36	0.753799798520829\\
0.365	0.753799798520829\\
0.37	0.753799798520829\\
0.375	0.753799798520829\\
0.38	0.753799798520829\\
0.385	0.753799798520829\\
0.39	0.762020767075249\\
0.395	0.762020767075249\\
0.4	0.762020767075249\\
0.405	0.762020767075249\\
0.41	0.762020767075249\\
0.415	0.762020767075249\\
0.42	0.762020767075249\\
0.425	0.762020767075249\\
0.43	0.762020767075249\\
0.435	0.769041506842049\\
0.44	0.769041506842049\\
0.445	0.769041506842049\\
0.45	0.769041506842049\\
0.455	0.769041506842049\\
0.46	0.769041506842049\\
0.465	0.769041506842049\\
0.47	0.769041506842049\\
0.475	0.769041506842049\\
0.48	0.769041506842049\\
0.485	0.769041506842049\\
0.49	0.769041506842049\\
0.495	0.769041506842049\\
0.5	0.769041506842049\\
};
\addlegendentry{Subset merging}

\addplot [color=mycolor3, line width=2.0pt, mark=o, mark options={scale=1.5, line width=0.8pt}]
  table[row sep=crcr]{%
0.005	0.00500000000000209\\
0.01	0.010000000000002\\
0.015	0.0150000000000018\\
0.02	0.0200000000000015\\
0.025	0.025000000000002\\
0.03	0.0300000000000015\\
0.035	0.0350000000000021\\
0.04	0.0400000000000019\\
0.045	0.0450000000000017\\
0.05	0.0500000000000017\\
0.055	0.0550000000000017\\
0.06	0.0600000000000017\\
0.065	0.0650000000000016\\
0.07	0.0700000000000018\\
0.075	0.0750000000000014\\
0.08	0.0800000000000018\\
0.085	0.0850000000000015\\
0.09	0.0900000000000018\\
0.095	0.0950000000000016\\
0.1	0.100000000000002\\
0.105	0.105000000000002\\
0.11	0.110000000000002\\
0.115	0.115000000000002\\
0.12	0.120000000000002\\
0.125	0.125000000000002\\
0.13	0.130000000000002\\
0.135	0.135000000000002\\
0.14	0.140000000000002\\
0.145	0.145000000000001\\
0.15	0.150000000000001\\
0.155	0.155000000000001\\
0.16	0.160000000000001\\
0.165	0.165000000000001\\
0.17	0.170000000000001\\
0.175	0.175000000000001\\
0.18	0.180000000000002\\
0.185	0.185000000000001\\
0.19	0.190000000000001\\
0.195	0.195000000000001\\
0.2	0.200000000000001\\
0.205	0.205000000000001\\
0.21	0.210000000000003\\
0.215	0.215000000000002\\
0.22	0.220000000000003\\
0.225	0.225000000000002\\
0.23	0.230000000000002\\
0.235	0.235000000000002\\
0.24	0.240000000000001\\
0.245	0.245000000000002\\
0.25	0.250000000000002\\
0.255	0.255000000000002\\
0.26	0.260000000000002\\
0.265	0.265000000000002\\
0.27	0.270000000000002\\
0.275	0.275000000000001\\
0.28	0.280000000000004\\
0.285	0.285000000000002\\
0.29	0.290000000000002\\
0.295	0.295000000000001\\
0.3	0.300000000000002\\
0.305	0.305000000000002\\
0.31	0.310000000000002\\
0.315	0.315000000000002\\
0.32	0.320000000000002\\
0.325	0.325000000000002\\
0.33	0.330000000000002\\
0.335	0.335000000000001\\
0.34	0.340000000000001\\
0.345	0.345000000000002\\
0.35	0.350000000000001\\
0.355	0.355000000000001\\
0.36	0.360000000000001\\
0.365	0.365000000000001\\
0.37	0.370000000000001\\
0.375	0.375000000000001\\
0.38	0.380000000000002\\
0.385	0.385000000000002\\
0.39	0.390000000000001\\
0.395	0.394999999999999\\
0.4	0.400000000000001\\
0.405	0.405000000000002\\
0.41	0.410000000000001\\
0.415	0.415000000000002\\
0.42	0.420000000000001\\
0.425	0.425000000000001\\
0.43	0.430000000000001\\
0.435	0.435000000000001\\
0.44	0.440000000000001\\
0.445	0.445000000000001\\
0.45	0.450000000000001\\
0.455	0.455000000000001\\
0.46	0.460000000000001\\
0.465	0.465000000000001\\
0.47	0.470000000000001\\
0.475	0.475000000000001\\
0.48	0.480000000000001\\
0.485	0.485\\
0.49	0.49\\
0.495	0.495000000000001\\
0.5	0.500000000000001\\
};
\addlegendentry{Max-lift privacy}

\end{axis}

\end{tikzpicture}%
}}
            \caption{\label{fig:funnelPUT} Privacy-utility tradeoff comparison between Algorithm \ref{alg:funnel}, subset merging \cite{2023Onthelift} (modified for semi-pointwise measure $\mathfrak{L}(y)$) and max-lift mechanism \cite{2021DataSanitize} for privacy funnel. } 
    \end{figure*}
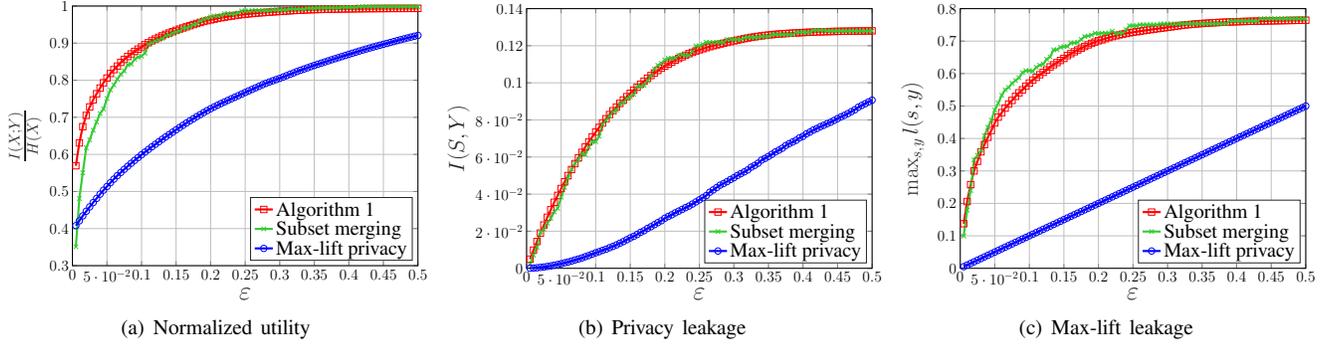

     \begin{figure*}[]
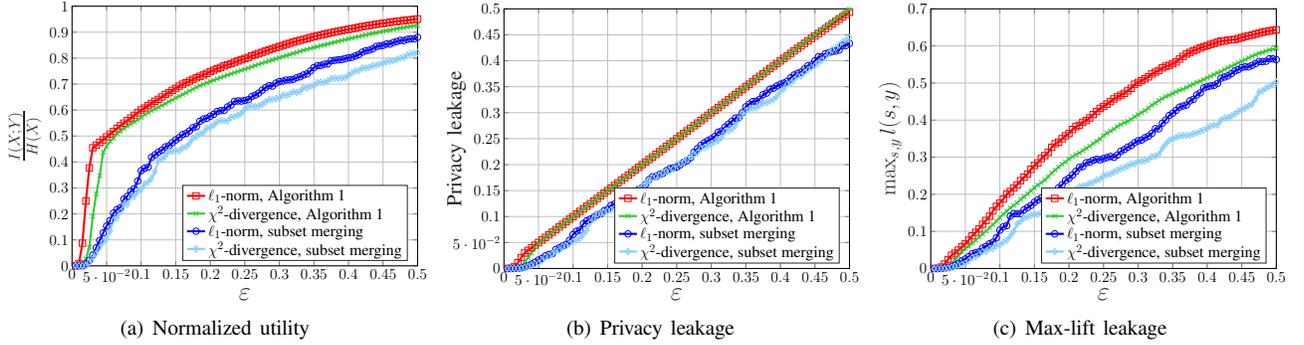

            \centering   
            \subfigure[\label{fig:chiutility}Normalized utility]{\scalebox{0.4}{\input{figures/4_UtilChi.tex}}}
            \subfigure[\label{fig:chileak} Privacy leakage]{\scalebox{0.4}{\input{figures/5_ISYChi.tex}}}
            \subfigure[\label{fig:chimax} Max-lift leakage]{\scalebox{0.4}{\input{figures/6_MaxChi.tex}}}
            \caption{\label{fig:chiPUT} Privacy-utility tradeoff comparison between Algorithm \ref{alg:funnel} and subset merging \cite{2023Onthelift} modified for $\ell_{1}$-norm and $\chi^2$-divergence. } 
     \end{figure*}

   \begin{figure*}[]
            \centering   
            \subfigure[\label{fig:zamaniutility} Utility]{\scalebox{0.4}{\definecolor{mycolor1}{rgb}{1,0,0}%
\definecolor{mycolor2}{rgb}{0.2, 0.8, 0.2}%
\definecolor{mycolor3}{rgb}{0,0,1}%
\definecolor{mycolor4}{rgb}{0.53, 0.81, 0.98}
\begin{tikzpicture}

\begin{axis}[%
width=4.521in,
    height=3.4in,
    scale only axis,
    xlabel style={font=\color{white!15!black}, at={(axis description cs:0.5,-0.01)}},
    every x tick label/.append style={font=\color{darkgray!60!black},font=\Large},
    xlabel={\Huge $\eps$},
    xmin=0,
    xmax=0.08,
    ymin=0,
    ymax=0.4,
    ylabel style={font=\color{white!15!black}, at={(axis description cs:-0.03,0.5)}},
    ylabel={\huge ${I(X;Y)}$},
    every y tick label/.append style={font=\color{darkgray!60!black},font=\Large},
    axis background/.style={fill=white},
    xmajorgrids,
    ymajorgrids,
legend style={at={(0.01,0.82)}, anchor=south west, legend cell align=left, align=left, draw=white!15!black,font=\fontsize{15}{20}\selectfont} ]
\addplot [color=mycolor1, line width=2.0pt,
mark=square, mark options={scale=1.5, line width=0.8pt}]
  table[row sep=crcr]{%
0.001	0\\
0.002	0.000189952270889449\\
0.003	0.000285741979881492\\
0.004	0.000774722356100577\\
0.005	0.00119024988015324\\
0.006	0.00173274796980498\\
0.007	0.00233273702427196\\
0.008	0.00310221574394065\\
0.009	0.00389091665654964\\
0.01	0.00483219961817376\\
0.011	0.00580688006164648\\
0.012	0.00699232046689487\\
0.013	0.00823900232872967\\
0.014	0.00950142772571834\\
0.015	0.0109483549208328\\
0.016	0.0124616199268939\\
0.017	0.0141145773756651\\
0.018	0.015760527504372\\
0.019	0.0176168475773945\\
0.02	0.0195339792206806\\
0.021	0.0216001264927714\\
0.022	0.0236372358692417\\
0.023	0.0260030203191193\\
0.024	0.0282413593848146\\
0.025	0.0307298066954468\\
0.026	0.033327228910329\\
0.027	0.0359796793596869\\
0.028	0.0387964835212218\\
0.029	0.0415473878813786\\
0.03	0.044582061645035\\
0.031	0.0476707874425292\\
0.032	0.0509323268089887\\
0.033	0.0541080296005722\\
0.034	0.0575955079454097\\
0.035	0.0611381868330515\\
0.036	0.0648619075965467\\
0.037	0.0687058711828287\\
0.038	0.0724405578656986\\
0.039	0.076696640760089\\
0.04	0.0806670475829195\\
0.041	0.0850013115502535\\
0.042	0.0892087199197124\\
0.043	0.0939927831907663\\
0.044	0.0984506961201263\\
0.045	0.103304643824862\\
0.046	0.108232471628119\\
0.047	0.113363359327347\\
0.048	0.118343908155339\\
0.049	0.123755342860627\\
0.05	0.129560645801321\\
0.051	0.134967531968801\\
0.052	0.140834020512774\\
0.053	0.146535321201694\\
0.054	0.152995422220818\\
0.055	0.159015762158266\\
0.056	0.165538515606415\\
0.057	0.17188894583234\\
0.058	0.179082803925457\\
0.059	0.18579636354023\\
0.06	0.193063416835843\\
0.061	0.200158648844557\\
0.062	0.208200666211673\\
0.063	0.216115746011084\\
0.064	0.223868447783595\\
0.065	0.232259774937052\\
0.066	0.240924677088461\\
0.067	0.249869198241212\\
0.068	0.258688096122818\\
0.069	0.268734758957148\\
0.07	0.278235728858752\\
0.071	0.288567719634782\\
0.072	0.298883945253093\\
0.073	0.310760983146458\\
0.074	0.322677130140623\\
0.075	0.334847355196187\\
0.076	0.348463954482713\\
0.077	0.363934884341162\\
0.078	0.382665918419661\\
};
\addlegendentry{$\chi^2$-divergence, Algorithm \ref{alg:funnel} }

\addplot [color=mycolor2, line width=1.0pt, mark=o, mark options={solid, mycolor2}]
  table[row sep=crcr]{%
0.001	4.87062419032542e-05\\
0.002	0.000194837421960547\\
0.003	0.000438430927015626\\
0.004	0.000779549147931727\\
0.005	0.00121827959918605\\
0.006	0.00175473508712982\\
0.007	0.00238905392774716\\
0.008	0.00312140021499999\\
0.009	0.00395196414110883\\
0.01	0.00488096237039296\\
0.011	0.00590863846858377\\
0.012	0.00703526338983775\\
0.013	0.0082611360239978\\
0.014	0.00958658380702214\\
0.015	0.0110119633978696\\
0.016	0.0125376614255659\\
0.017	0.0141640953106246\\
0.018	0.0158917141655074\\
0.019	0.0177209997793626\\
0.02	0.0196524676928997\\
0.021	0.0216866683699416\\
0.022	0.0238241884729577\\
0.023	0.0260656522507406\\
0.024	0.0284117230473398\\
0.025	0.0308631049424441\\
0.026	0.0334205445346246\\
0.027	0.0360848328802158\\
0.028	0.0388568076021754\\
0.029	0.0417373551850357\\
0.03	0.0447274134740862\\
0.031	0.0478279743992324\\
0.032	0.0510400869466441\\
0.033	0.0543648604043564\\
0.034	0.0578034679115203\\
0.035	0.0613571503450948\\
0.036	0.0650272205825175\\
0.037	0.0688150681844439\\
0.038	0.0727221645481338\\
0.039	0.0767500685896797\\
0.04	0.0809004330222755\\
0.041	0.0851750113083571\\
0.042	0.0895756653761132\\
0.043	0.0941043742059877\\
0.044	0.0987632434109318\\
0.045	0.103554515956055\\
0.046	0.108480584189849\\
0.047	0.113544003391493\\
0.048	0.118747507078382\\
0.049	0.124094024366935\\
0.05	0.129586699740429\\
0.051	0.135228915653545\\
0.052	0.141024318498891\\
0.053	0.146976848582188\\
0.054	0.153090774908231\\
0.055	0.159370735780569\\
0.056	0.165821786479943\\
0.057	0.172449455632178\\
0.058	0.179259812337292\\
0.059	0.186259546754551\\
0.06	0.193456067691148\\
0.061	0.200857621927998\\
0.062	0.208473441691766\\
0.063	0.216313929093897\\
0.064	0.224390889899283\\
0.065	0.232717834309\\
0.066	0.241310370645239\\
0.067	0.250186730848005\\
0.068	0.259368488066391\\
0.069	0.268881563103387\\
0.07	0.278757681646286\\
0.071	0.289036567230383\\
0.072	0.299769403162222\\
0.073	0.31102464156098\\
0.074	0.322898571076466\\
0.075	0.335536838159218\\
0.076	0.349186465987765\\
0.077	0.36436632206785\\
0.078	0.383425122666222\\
};
\addlegendentry{$\chi^2$-divergence, theoretical \cite{2021StrongChi2}}

\end{axis}
\end{tikzpicture}%
}}
            \subfigure[\label{fig:zamanileak} Privacy leakage]{\scalebox{0.4}{\definecolor{mycolor1}{rgb}{1,0,0}%
\definecolor{mycolor2}{rgb}{0.2, 0.8, 0.2}%
\definecolor{mycolor3}{rgb}{0,0,1}%
\definecolor{mycolor4}{rgb}{0.53, 0.81, 0.98}
\begin{tikzpicture}

\begin{axis}[%
width=4.521in,
height=3.4in,
scale only axis,
xlabel style={font=\color{white!15!black}},
every x tick label/.append style={font=\color{darkgray!60!black},font=\Large},
xlabel={\Huge $\eps$},
xmin=0,
xmax=0.08,
ymin=0,
ymax=0.007,
ylabel style={font=\color{white!15!black}, at={(axis description cs:-0.01,0.5)}},
ylabel={\huge $\max_{y}\chi^2(y) $},
every y tick label/.append style={font=\color{darkgray!60!black},font=\Large},
axis background/.style={fill=white},
xmajorgrids,
ymajorgrids,
legend style={at={(0.01,0.82)}, anchor=south west, legend cell align=left, align=left, draw=white!15!black,font=\fontsize{15}{20}\selectfont} ]
\addplot [color=mycolor1, line width=2.0pt,
mark=square, mark options={scale=1.5, line width=0.8pt}]
  table[row sep=crcr]{%
0.001	0\\
0.002	3.95689655172562e-06\\
0.003	8.64882352940709e-06\\
0.004	1.59727450980433e-05\\
0.005	2.47694117647024e-05\\
0.006	3.56120689655075e-05\\
0.007	4.81286274509786e-05\\
0.008	6.38909803921435e-05\\
0.009	8.0523333333339e-05\\
0.01	9.90776470588097e-05\\
0.011	0.000119553921568637\\
0.012	0.000143754705882349\\
0.013	0.000168897931034503\\
0.014	0.000194612745098038\\
0.015	0.000224558275862105\\
0.016	0.000255563921568643\\
0.017	0.000288132413793089\\
0.018	0.000322093333333353\\
0.019	0.000359620344827587\\
0.02	0.000399318627450937\\
0.021	0.000439458039215689\\
0.022	0.000481519411764665\\
0.023	0.000528965686274533\\
0.024	0.000575018823529384\\
0.025	0.000622993921568657\\
0.026	0.000675591724137895\\
0.027	0.000728775686274569\\
0.028	0.000782955517241374\\
0.029	0.000838475294117603\\
0.03	0.000898233103448296\\
0.031	0.000960531176470564\\
0.032	0.00102225568627456\\
0.033	0.00108590215686272\\
0.034	0.00115252965517236\\
0.035	0.00122423215686276\\
0.036	0.00129379235294112\\
0.037	0.00136892793103446\\
0.038	0.00143867862745095\\
0.039	0.00152008137931036\\
0.04	0.00159727450980391\\
0.041	0.0016791486206897\\
0.042	0.00175783215686276\\
0.043	0.00184747058823524\\
0.044	0.0019327021568628\\
0.045	0.00202102448275857\\
0.046	0.00211586274509813\\
0.047	0.00220700803921568\\
0.048	0.00230007529411756\\
0.049	0.00239506450980394\\
0.05	0.00249951000000005\\
0.051	0.00259849098039222\\
0.052	0.00270236689655183\\
0.053	0.0028022188235295\\
0.054	0.00291510274509805\\
0.055	0.00302191941176461\\
0.056	0.00313182206896549\\
0.057	0.0032413186274509\\
0.058	0.00336264098039224\\
0.059	0.00347729333333331\\
0.06	0.00359386764705868\\
0.061	0.00371236392156862\\
0.062	0.00384212470588226\\
0.063	0.00396788793103437\\
0.064	0.00408902274509798\\
0.065	0.00422244827586201\\
0.066	0.00435355392156861\\
0.067	0.0044849224137931\\
0.068	0.00461612333333334\\
0.069	0.00476069117647049\\
0.07	0.00489692862745106\\
0.071	0.00503508803921563\\
0.072	0.0051751694117648\\
0.073	0.00532817568627451\\
0.074	0.00547571172413785\\
0.075	0.00561824392156865\\
0.076	0.00577406172413791\\
0.077	0.00592760568627461\\
0.078	0.00608032551724144\\
};
\addlegendentry{$\chi^2$-divergence, Algorithm \ref{alg:funnel} }

\addplot [color=mycolor2, line width=1.0pt, mark=o, mark options={solid, mycolor2}]
  table[row sep=crcr]{%
0.001	9.99991208762804e-07\\
0.002	3.99996483505084e-06\\
0.003	8.99992087886423e-06\\
0.004	1.59998593402028e-05\\
0.005	2.49997802190675e-05\\
0.006	3.59996835154567e-05\\
0.007	4.89995692293714e-05\\
0.008	6.39994373608101e-05\\
0.009	8.09992879097777e-05\\
0.01	9.99991208762684e-05\\
0.011	0.000120998936260284\\
0.012	0.000143998734061825\\
0.013	0.000168998514280893\\
0.014	0.000195998276917488\\
0.015	0.000224998021971605\\
0.016	0.000255997749443247\\
0.017	0.000288997459332415\\
0.018	0.000323997151639108\\
0.019	0.000360996826363329\\
0.02	0.000399996483505076\\
0.021	0.000440996123064342\\
0.022	0.000483995745041139\\
0.023	0.000528995349435463\\
0.024	0.000575994936247305\\
0.025	0.000624994505476679\\
0.026	0.00067599405712357\\
0.027	0.000728993591187991\\
0.028	0.000783993107669941\\
0.029	0.000840992606569413\\
0.03	0.00089999208788641\\
0.031	0.000960991551620936\\
0.032	0.00102399099777298\\
0.033	0.00108899042634256\\
0.034	0.00115598983732966\\
0.035	0.00122498923073428\\
0.036	0.00129598860655643\\
0.037	0.00136898796479611\\
0.038	0.00144398730545331\\
0.039	0.00152098662852804\\
0.04	0.00159998593402029\\
0.041	0.00168098522193005\\
0.042	0.00176398449225737\\
0.043	0.00184898374500219\\
0.044	0.00193598298016455\\
0.045	0.00202498219774443\\
0.046	0.00211598139774183\\
0.047	0.00220898058015676\\
0.048	0.00230397974498922\\
0.049	0.0024009788922392\\
0.05	0.0024999780219067\\
0.051	0.00260097713399173\\
0.052	0.00270397622849429\\
0.053	0.00280897530541437\\
0.054	0.00291597436475197\\
0.055	0.0030249734065071\\
0.056	0.00313597243067976\\
0.057	0.00324897143726994\\
0.058	0.00336397042627765\\
0.059	0.00348096939770288\\
0.06	0.00359996835154565\\
0.061	0.00372096728780593\\
0.062	0.00384396620648374\\
0.063	0.00396896510757907\\
0.064	0.00409596399109194\\
0.065	0.00422496285702231\\
0.066	0.00435596170537024\\
0.067	0.00448896053613566\\
0.068	0.00462395934931863\\
0.069	0.00476095814491911\\
0.07	0.00489995692293712\\
0.071	0.00504095568337267\\
0.072	0.00518395442622572\\
0.073	0.00532895315149631\\
0.074	0.00547595185918442\\
0.075	0.00562495054929008\\
0.076	0.00577594922181323\\
0.077	0.00592894787675393\\
0.078	0.00608394651411214\\
};
\addlegendentry{$\chi^2$-divergence, theoretical \cite{2021StrongChi2}}

\end{axis}

\end{tikzpicture}%
}}
            \subfigure[\label{fig:zamanimax} Max-lift leakage]{\scalebox{0.4}{\definecolor{mycolor1}{rgb}{1,0,0}%
\definecolor{mycolor2}{rgb}{0.2, 0.8, 0.2}%
\definecolor{mycolor3}{rgb}{0,0,1}%
\definecolor{mycolor4}{rgb}{0.53, 0.81, 0.98}
\begin{tikzpicture}

\begin{axis}[%
width=4.521in,
height=3.4in,
scale only axis,
xlabel style={font=\color{white!15!black}},
every x tick label/.append style={font=\color{darkgray!60!black},font=\Large},
xlabel={\Huge $\eps$},
xlabel style={font=\color{white!15!black}},
xmin=0,
xmax=0.08,
ymin=0,
ymax=0.1,
ylabel style={font=\color{white!15!black}, at={(axis description cs:-0.09 ,0.5)}},
ylabel={\huge $\max_{s,y}l(s,y)$},
every y tick label/.append style={font=\color{darkgray!60!black},font=\Large},
axis background/.style={fill=white},
xmajorgrids,
ymajorgrids,
legend style={at={(0.01,0.82)}, anchor=south west, legend cell align=left, align=left, draw=white!15!black,font=\fontsize{15}{20}\selectfont} ]
\addplot [color=mycolor1, line width=2.0pt,
mark=square, mark options={scale=1.5, line width=0.8pt}]
  table[row sep=crcr]{%
0.001	0\\
0.002	0.00259662584726591\\
0.003	0.00389241471534253\\
0.004	0.00528600442923854\\
0.005	0.00657831536012251\\
0.006	0.0078689583786963\\
0.007	0.0091579377847655\\
0.008	0.0105442138756702\\
0.009	0.0118297517535777\\
0.01	0.0131136391453821\\
0.011	0.0143958802837328\\
0.012	0.0157749191153618\\
0.013	0.0170537545658287\\
0.014	0.0183309566847232\\
0.015	0.0196065296389168\\
0.016	0.0209784063851923\\
0.017	0.0222506089348191\\
0.018	0.0235211950413465\\
0.019	0.0247901688072183\\
0.02	0.0261549574768499\\
0.021	0.0274205955759924\\
0.022	0.0286846338599076\\
0.023	0.0300441213483773\\
0.024	0.0313048498284117\\
0.025	0.0325639908732634\\
0.026	0.03382154847551\\
0.027	0.0351740750076554\\
0.028	0.0364283566135516\\
0.029	0.0376810669676874\\
0.03	0.038932210001788\\
0.031	0.0402778464985697\\
0.032	0.0415257468284994\\
0.033	0.0427720918438686\\
0.034	0.0440168854167728\\
0.035	0.0453557017207974\\
0.036	0.0465972853767812\\
0.037	0.0478373294141611\\
0.038	0.049075837646592\\
0.039	0.0504079025438466\\
0.04	0.0516432331518382\\
0.041	0.0528770396007655\\
0.042	0.054109325647032\\
0.043	0.0554347068880999\\
0.044	0.0566638471175426\\
0.045	0.057891478415777\\
0.046	0.0592118596318473\\
0.047	0.0604363688038305\\
0.048	0.0616593803867265\\
0.049	0.062880898039201\\
0.05	0.064194712041996\\
0.051	0.065413138548233\\
0.052	0.0666300822977626\\
0.053	0.0678455468950683\\
0.054	0.0691528586883167\\
0.055	0.0703652626605626\\
0.056	0.0715761984892222\\
0.057	0.0727856697256468\\
0.058	0.074086543352699\\
0.059	0.0752929840354311\\
0.06	0.0764979709727233\\
0.061	0.077701507663852\\
0.062	0.0789960062264971\\
0.063	0.080196541994277\\
0.064	0.0813956382039932\\
0.065	0.082593298303854\\
0.066	0.0838814839807018\\
0.067	0.0850761723551176\\
0.068	0.0862694351521757\\
0.069	0.0875528970078606\\
0.07	0.0887432098343892\\
0.071	0.0899321075006064\\
0.072	0.0911195933674846\\
0.073	0.0923968492659674\\
0.074	0.0935814136215294\\
0.075	0.0947645764444233\\
0.076	0.095946341047208\\
0.077	0.0972174507153599\\
0.078	0.0983963218561927\\
};
\addlegendentry{$\chi^2$-divergence, Algorithm \ref{alg:funnel} }

\addplot [color=mycolor2, line width=1.0pt, mark=o, mark options={solid, mycolor2}]
  table[row sep=crcr]{%
0.001	0.00132524561192053\\
0.002	0.00264873727206945\\
0.003	0.00397047961699322\\
0.004	0.00529047726487647\\
0.005	0.00660873481564129\\
0.006	0.00792525685104143\\
0.007	0.009240047934758\\
0.008	0.0105531126124956\\
0.009	0.0118644554120755\\
0.01	0.01317408084353\\
0.011	0.0144819933991952\\
0.012	0.0157881975538029\\
0.013	0.0170926977645736\\
0.014	0.0183954984713074\\
0.015	0.0196966040964732\\
0.016	0.0209960190453016\\
0.017	0.0222937477058725\\
0.018	0.023589794449203\\
0.019	0.0248841636293386\\
0.02	0.0261768595834382\\
0.021	0.0274678866318638\\
0.022	0.028757249078265\\
0.023	0.0300449512096665\\
0.024	0.0313309972965516\\
0.025	0.0326153915929509\\
0.026	0.0338981383365224\\
0.027	0.0351792417486392\\
0.028	0.0364587060344699\\
0.029	0.0377365353830626\\
0.03	0.0390127339674281\\
0.031	0.04028730594462\\
0.032	0.0415602554558173\\
0.033	0.0428315866264043\\
0.034	0.044101303566051\\
0.035	0.0453694103687926\\
0.036	0.0466359111131098\\
0.037	0.0479008098620064\\
0.038	0.0491641106630873\\
0.039	0.0504258175486371\\
0.04	0.0516859345356974\\
0.041	0.0529444656261419\\
0.042	0.0542014148067553\\
0.043	0.0554567860493062\\
0.044	0.0567105833106249\\
0.045	0.0579628105326767\\
0.046	0.0592134716426366\\
0.047	0.0604625705529639\\
0.048	0.0617101111614738\\
0.049	0.0629560973514129\\
0.05	0.0642005329915294\\
0.051	0.0654434219361474\\
0.052	0.0666847680252367\\
0.053	0.0679245750844841\\
0.054	0.0691628469253655\\
0.055	0.0703995873452153\\
0.056	0.0716348001272956\\
0.057	0.0728684890408675\\
0.058	0.074100657841258\\
0.059	0.0753313102699304\\
0.06	0.076560450054551\\
0.061	0.0777880809090575\\
0.062	0.0790142065337266\\
0.063	0.0802388306152414\\
0.064	0.0814619568267556\\
0.065	0.0826835888279623\\
0.066	0.0839037302651581\\
0.067	0.0851223847713092\\
0.068	0.0863395559661157\\
0.069	0.0875552474560761\\
0.07	0.0887694628345525\\
0.071	0.089982205681833\\
0.072	0.0911934795651955\\
0.073	0.092403288038971\\
0.074	0.0936116346446054\\
0.075	0.0948185229107229\\
0.076	0.0960239563531867\\
0.077	0.0972279384751611\\
0.078	0.0984304727671723\\
};
\addlegendentry{$\chi^2$-divergence, theoretical \cite{2021StrongChi2}}

\end{axis}
\end{tikzpicture}%
}}
            \caption{\label{fig:zamaniPUT} Privacy-utility tradeoff comparison between Algorithm \ref{alg:funnel} and theoretical framework in \cite{2021StrongChi2} for and $\chi^2$-divergence.} 
    \end{figure*}
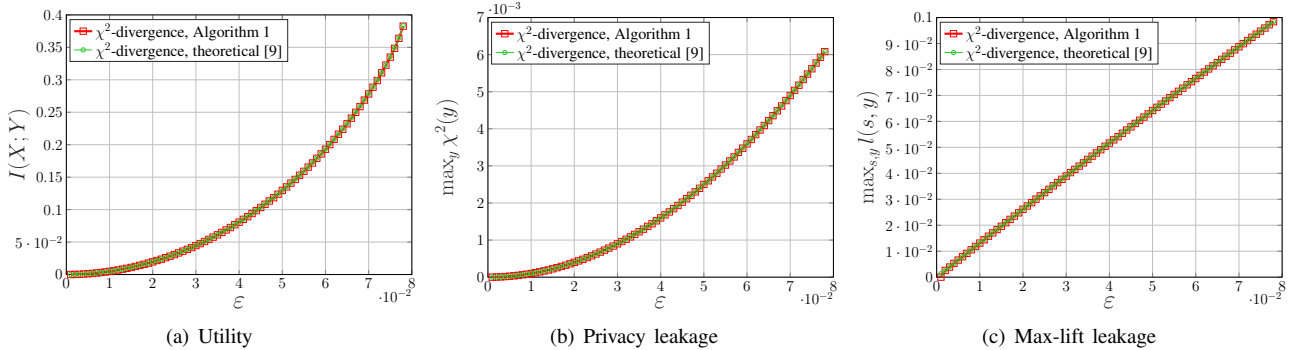

    In this section, we demonstrate the performance of Algorithm \ref{alg:funnel} and compare it with previous results in \cite{2023Onthelift} and \cite{2021DataSanitize} through numerical evaluation.
    For this purpose, we have generated $100$ distributions as $\PSX$, where $|\mathcal{S}|=4$ and $|\mathcal{X}|=7$, and the presented results have been averaged over these distributions. For all mentioned privacy measures, we have depicted the normalized utility $\frac{I(X;Y)}{H(X)}$, privacy leakage and max-lift leakage versus $\eps$, where $\mathcal{E}=\{0.0025,0.005,0.0075,\cdots,0.5\}$, $\eps_{|\mathcal{E}|+1}=1$, $n_{i}=5~~\text{for}~~ 1\leq i \leq |\mathcal{E}|-1$, $n_{|\mathcal{E}|}=500$ and $\delta=0.05$.

    The main reference for comparison is the subset merging method proposed in \cite{2023Onthelift}. This method has been designed based on the watchdog privacy mechanism \cite{2019Watchdog,2020PropertiesWatchdog}, where the $\X$ alphabet is divided into high-risk and low-risk symbols according to the privacy measure and privacy budget, and only high-risk ones are randomized. Subset merging enhances the utility of the watchdog mechanism, has a much lower complexity than the optimal max-lift mechanism, and is a flexible privacy mechanism that can be applied to pointwise and semi-pointwise measures.
    
    Fig. \ref{fig:funnelPUT} shows the PUT for privacy funnel given by Algorithm \ref{alg:funnel}, subset merging method \cite{2023Onthelift} and optimal max-lift mechanism \cite{2021DataSanitize}. To have a fair comparison, both Algorithm \ref{alg:funnel} and subset merging method have been applied for semi-pointwise measure $\mathfrak{L}(y)$. Accordingly, the results here differ from \cite{2023Onthelift}, where the privacy measure was local information privacy (both max-lift and min-lift). 
    However, for the optimal max-lift mechanism, we kept lift as the privacy measure to show the benefits of applying semi-pointwise measure $\mathfrak{L}(y)$ instead of max-lift as a (fully) pointwise measure. Furthermore, these results are also different from \cite{2021DataSanitize} since, in that paper, both max-lift and min-lift were bounded, while here, we only bound max-lift.

    Fig. \ref{fig:utilityfunnel} compares the utility of these three methods. By comparing the blue curve with red and green ones, it is readily confirmed that using semi-pointwise measure $\mathfrak{L}(y)$ instead of the max-lift can boost utility for the privacy funnel.      
    Additionally, Algorithm \ref{alg:funnel} performs better than subset merging for small values of epsilon $\eps$ (the high privacy regime) as the effective range for privacy.  It enhances utility for $\eps$ ranging from $0.0025$ to $0.17$, with utility improving from $67\%$ at $\eps = 0.0025$ to less than $2\%$ at $0.105 \leq \eps \leq 0.17$.  When $\eps$ passes $0.17$, subset merging results in a little bit better utility, however, such flattened curve involving very high utility values indicate that in this range of $\eps$, there is no meaningful privacy protection. The reason is that in this range of $\eps$, subset merging determines one or two high-risk symbols and releases all the other ones without randomization, which causes higher utility but also risks full disclosure of those symbols that are deemed low-risk.

    Interestingly, this enhancement in utility happens with almost the same amount of privacy leakage $I(S;Y)$ as shown in Fig. \ref{fig:funnelleak}. 
    In Fig. \ref{fig:funnelmax}, we can see that max-lift leakage in the effective range of $\eps$ for subset merging is higher for most values of $\eps$ than the max-lift leakage of Algorithm \ref{alg:funnel}, indicating that Algorithm \ref{alg:funnel} enhances utility with lower max-lift leakage that indicates the efficiency of {this approach.}

    Fig. \ref{fig:chiPUT} demonstrates the PUT for $\ell_{1}$-norm and $\chi^2$-divergence respectively. We emphasize that Algorithm \ref{alg:funnel} remains essentially the same for $\ell_{1}$-norm and $\chi^2$-divergence privacy measures and the only difference is the use of these metrics instead of semi-pointwise measure $\mathfrak{L}(y)$ for privacy. 
    Again, we compared the output of Algorithm \ref{alg:funnel} with subset merging for these lift-based measures. 
    Fig. \ref{fig:chiutility} displays that Algorithm \ref{alg:funnel} greatly improves utility for very small values of $\eps$. 
    As the privacy budget increases, the subset merging utility gets closer to the Algorithm \ref{alg:funnel} results. 
    The reason for better performance for these measures compared to the privacy funnel is their convexity w.r.t the lift. This property causes Algorithm \ref{alg:funnel} to estimate the optimal solution. 
    Fig. \ref{fig:chileak} shows the privacy leakages which are $\max_{y}\ell_{1}(y)$ and $\sqrt{\max_{y}\chi^2(y)}$, where we took a square root for $\chi^{2}(y)$ since it has been bounded by $\eps^{2}$. This figure shows that Algorithm \ref{alg:funnel} is capable of meeting extreme values of privacy measures since both $\ell_{1}$-norm and $\chi^{2}$-divergence achieved the value of $\eps$. It also confirms Corollary \ref{corl:zamani} as $\ell_{1}$-norm results in a little bit higher utility than $\chi^{2}$-divergence with the same privacy budget. Moreover, Fig. \ref{fig:chimax} clarifies the reason behind this phenomenon as the maximum lift leakage for the $\ell_{1}$-norm is higher than that for the $\chi^{2}$-divergence.

    \subsection{Validation with theoretical framework in \cite{2021StrongChi2}}
    
    Here, we compare our method with the theoretical framework in \cite{2021StrongChi2}, where the vertices of the polytope and optimal utility for $\chi^2$-divergence have been estimated for small privacy budgets. For completeness, let us recall necessary numerical quantities from \cite[Example 1]{2021StrongChi2}.
    
    \begin{example}[\cite{2021StrongChi2}]\label{exmp:zamani}
        Consider the matrix $ \small P_{S|X}=\left[\begin{array}{cc}0.25 & 0.4 \\ 0.75 & 0.6 \end{array}\right]$ and $P_X$  as $\left[0.25, 0.75 \right]^T$. Then we have  $P_S = P_{S|X} P_X= [0.3625,0.6375]^T.$ Finally, the optimal $\PXgY$ is given as
        \begin{align}
            P_{X|Y=0} & =[0.25-3.2048 \cdot \eps, 0.75+3.2048 \cdot \eps]^T, \\
            P_{X|Y=1} & =[0.25+3.2048 \cdot \eps, 0.75-3.2048 \cdot \eps]^T .
        \end{align}
    Note that the approximation is valid for $\eps \ll 0.078.$
    \end{example}
    
    In Fig. \ref{fig:zamaniPUT}, we have applied Algorithm \ref{alg:funnel} to Example \ref{exmp:zamani} where we can observe a perfect match of our heuristic method to the theoretical framework. 
    Note that the method in \cite{2021StrongChi2} requires invertible  $\PSgX$ and very small $\epsilon$, while Algorithm \ref{alg:funnel} is simpler and has no such requirements.

\bibliographystyle{IEEEtran}
\bibliography{IEEEabrv.bib, BIB}
\end{document}